\documentclass[12pt]{article}

\usepackage{scicite}
\usepackage{times,deluxetable}
\usepackage{aas_macros}
\usepackage{upgreek,float}

\newcommand{\e}{\epsilon}
\newcommand{\g}{\gamma}

\newcommand{\psim}{\lower.5ex\hbox{$\; \buildrel \propto \over\sim \;$}}
\newcommand{\lbar}{\lower.0ex\hbox{$\; \buildrel
{\lower0.0ex \hbox{-}} \over\lambda  \;$}}

\newcommand{\mic}{{\rm $\upmu$m}}
\newcommand{\gsim}{\gtrsim}

\newcommand{\Hz}{\mathrm{Hz}}

\newcommand{\tgg}{$\tau_{\gamma\gamma}$}

\usepackage{amsmath,graphicx}
\usepackage{lineno}

\date{}

\usepackage{calrsfs,euscript,mathrsfs,amssymb}

\newenvironment{sciabstract}{%
\begin{quote} \bf}
{\end{quote}}

\title{A $\gamma$-ray determination of the Universe's star-formation history}

\begin{document}

\maketitle





%
%
%
%


\begin{center}
The Fermi-LAT Collaboration$^{*,\dagger}$ 
\end{center}

$*$ The list of authors and affiliations can be found at the beginning of
the supplementary material\\

$\dagger$ Corresponding Authors: majello@g.clemson.edu, helgason@hi.is,
  vpaliya@g.clemson.edu,\\
  justin.finke@nrl.navy.mil, abhishd@g.clemson.edu, alberto@gae.ucm.es

\begin{sciabstract}

The light emitted by all galaxies over the history of the Universe produces the
extragalactic background light (EBL) at
ultraviolet, optical, and infrared wavelengths. The EBL is a
 source of
opacity for $\gamma$ rays via photon-photon
interactions, leaving an 
imprint in the spectra of distant $\gamma$-ray
sources. We measure this
attenuation using {739} active galaxies and one gamma-ray
burst detected by the {\it Fermi} Large
Area Telescope. This allows us to reconstruct the
evolution of the EBL 
 and determine the star-formation history of the Universe over 90\% of cosmic
time. Our star-formation history is consistent
with independent measurements from galaxy surveys, peaking at
redshift $z\sim2$.
Upper limits of the EBL at the epoch of re-ionization suggest a turnover
 in the abundance of faint galaxies
at $z\sim 6$.
\end{sciabstract}

%
%
Stars 
 produce the bulk of the optical light in the Universe and synthesize most of the elements found
in galaxies.  
The cosmic star-formation history (SFH), i.e. the stellar birth rate as a
function of the Universe's age, summarizes the history of stellar
formation since the Big Bang \cite{Madau2014}.
The rate of star formation is commonly estimated by measuring direct emission of
light from massive short-lived stars, typically in the ultraviolet (UV) and (or) by
detecting the reprocessed radiation from dusty star-forming regions
in the infrared (IR). The conversion from the UV light emitted by  a minority
of stars to the 
stellar mass formed per year relies on assumptions about the mass
distribution of the newly formed stellar population (the initial
mass function, IMF), the element enrichment history of the interstellar
medium, and  obscuration by dust. Such
estimates of the SFH rely on the detection of many individual galaxies
in deep surveys \cite{Grogin2011,Illingworth2013,Lotz2017}. Because not even the most powerful telescope can detect
all the galaxies in a representative field, one of the
largest sources of uncertainty in the SFH is estimating the amount of light from undetected
galaxies, and the star formation associated with them. This
difficulty becomes particularly relevant within the first billion
years after the Big Bang when a large population of
faint, still undetected, galaxies existed
\cite{McLure13}.   
These galaxies are expected to drive the re-ionization of the
Universe: the period when energetic UV photons from young
stars escaped into intergalactic space and ionized the neutral
hydrogen of the intergalactic medium. Similarly,  recent (i.e. within
one billion years from the present age) star formation measured using space-borne UV observatories
is based on surveys extending over small solid angles \cite{Schiminovich2005}, and are therefore
subject to density fluctuations in the large-scale structure, an
effect known as cosmic variance.

Observational estimates of the SFH 
are sufficiently
uncertain that measurements with multiple independent methodologies are desirable.
Starlight
that escapes galaxies is almost never destroyed and becomes part
of the extragalactic background light (EBL), the total light
accumulated by all sources over the lifetime of the Universe
\cite{hauser2001,kashlinsky2005,dwek13}.  
While
extremely important, accurate measurements of this diffuse all-sky
background at UV to IR wavelengths, and particularly its build-up
over time, have only just become possible \cite{Andrews17}.

We present an alternative approach to measure the SFH based on the
attenuation that the EBL produces in the $\gamma$-ray spectra of
distant sources.  $\gamma$ rays with sufficient energy can
annihilate when they collide with EBL photons and produce 
  electron-positron pairs (i.e. the reaction $\gamma\gamma \rightarrow e^+e^-$),
  effectively being absorbed as a result of the
interaction \cite{nikishov62}. Above a given threshold
energy, the attenuation experienced by every $\gamma$-ray source at
 a similar distance depends on the number density of the EBL target
photons integrated along the line of sight; observations of
$\gamma$-ray sources at different distances (as measured by the sources’ redshifts)
can be used to measure the
density of EBL photons at different cosmic times.

We analyze $\gamma$-ray photons detected by the Large
Area Telescope (LAT) instrument on the {\it Fermi} Gamma-ray Space Telescope, over 9 years of
operations. Our sample of suitable objects for this analysis consists
of {739} blazars, galaxies
hosting a super-massive black hole with a relativistic jet pointed at
a small angle to the
line of sight. The distances of these blazars correspond to  lookback times of 0.2-11.6
billion years according to the standard cosmological
model \cite{note1}.
 We perform a likelihood analysis to find the
EBL attenuation experienced by all blazars whilst simultaneously optimizing
the spectral parameters independently for  each blazar \cite{ebl12}. This is
accomplished individually for each source, by defining a region of interest that
comprises all $\gamma$ rays detected within 15$^{\circ}$ of the source
position and creating a sky model that includes all sources of
$\gamma$ rays in the field. The parameters of the sky model are then
optimized by a maximum likelihood method.
For every blazar, the fitting is
performed below an energy at which the EBL attenuation is negligible and
thus yields a measurement of the intrinsic (i.e., unabsorbed) blazar
spectrum. The intrinsic spectra are described using simple empirical
functions \cite{som} and extrapolated
to higher energy, where the $\gamma$ rays are expected to be attenuated
by the EBL.

Potential EBL absorption is added to the fitted spectra as follows:
\begin{equation}
\left(\frac{{\rm d} N}{{\rm d} E} \right)_{\rm obs} = \left( \frac{{\rm d}N}{{\rm d} E} \right)_{\rm int}
\times {\rm e}^{-b \cdot \tau_{\gamma\gamma}(E,z)}
\label{eq:obs}
\end{equation}
where $\left(\frac{{\rm d}N}{{\rm d}E} \right)_{\rm obs}$ and $ \left( \frac{{\rm d}N}{{\rm d}E}
\right)_{\rm int}$ are the observed and intrinsic blazar spectra
respectively, $\tau_{\gamma\gamma}(E,z)$ is the EBL optical depth as estimated from
models (at a given energy $E$ and redshift $z$) and $b$ is a free parameter. The data from all blazars are
combined  to yield the best-fitting value of $b$ for each model. A
value of $b=0$ implies no EBL attenuation is present in the spectra
of blazars, while $b\approx1$ implies an attenuation compatible with
the model prediction. Twelve of  the most recent models that
predict the EBL attenuation up to a redshift of $z=3.1$ have been
tested in this work.
 We detect the
attenuation due to the EBL in the spectra of blazars at 
$\gtrsim16$ standard deviations ($\sigma$) for all models tested (see Table~S2).

\begin{figure*}[ht!]
  \begin{center}
  \begin{tabular}{c}
  	 \includegraphics[scale=0.45]{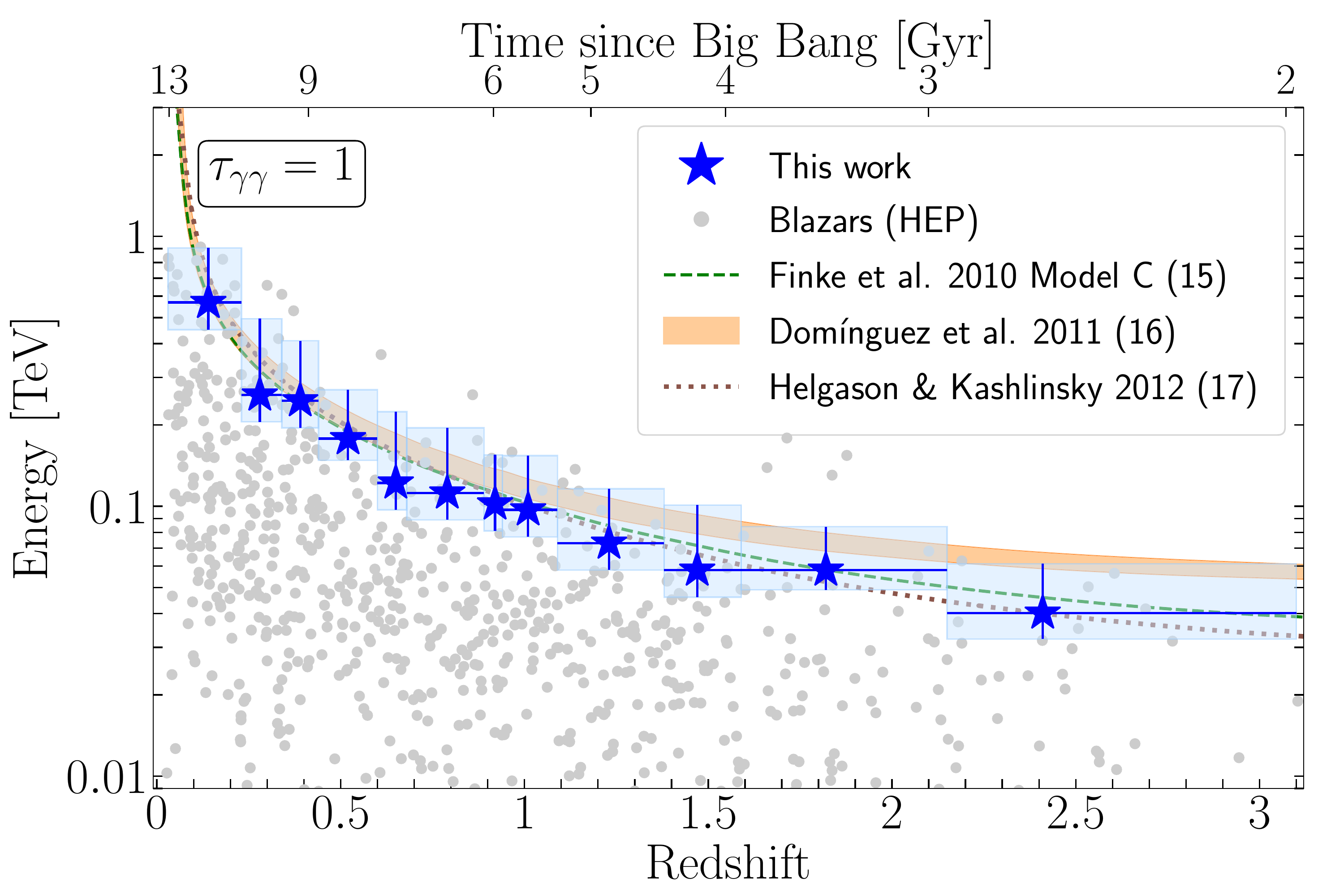} 
\end{tabular}
  \end{center}
  \caption{{\bf The cosmic $\gamma$-ray horizon.} Measurement of the
    cosmic $\gamma$-ray horizon ($\tau_{\gamma\gamma} = 1$, i.e. the point after
    which the Universe becomes opaque to $\gamma$ rays)
    as a
    function of redshift (blue stars and boxes, the latter
    representing the redshift bin size and the uncertainty on the energy) compared with predictions from
    three different EBL models
    \cite{finke10,dominguez11,helgason12}. The gray  points show
    the highest-energy photon (HEP) detected from each blazar considered in this
    work.
\label{fig:horizon}}
\end{figure*}

\begin{figure}
\centering
  \includegraphics[width=1\textwidth]{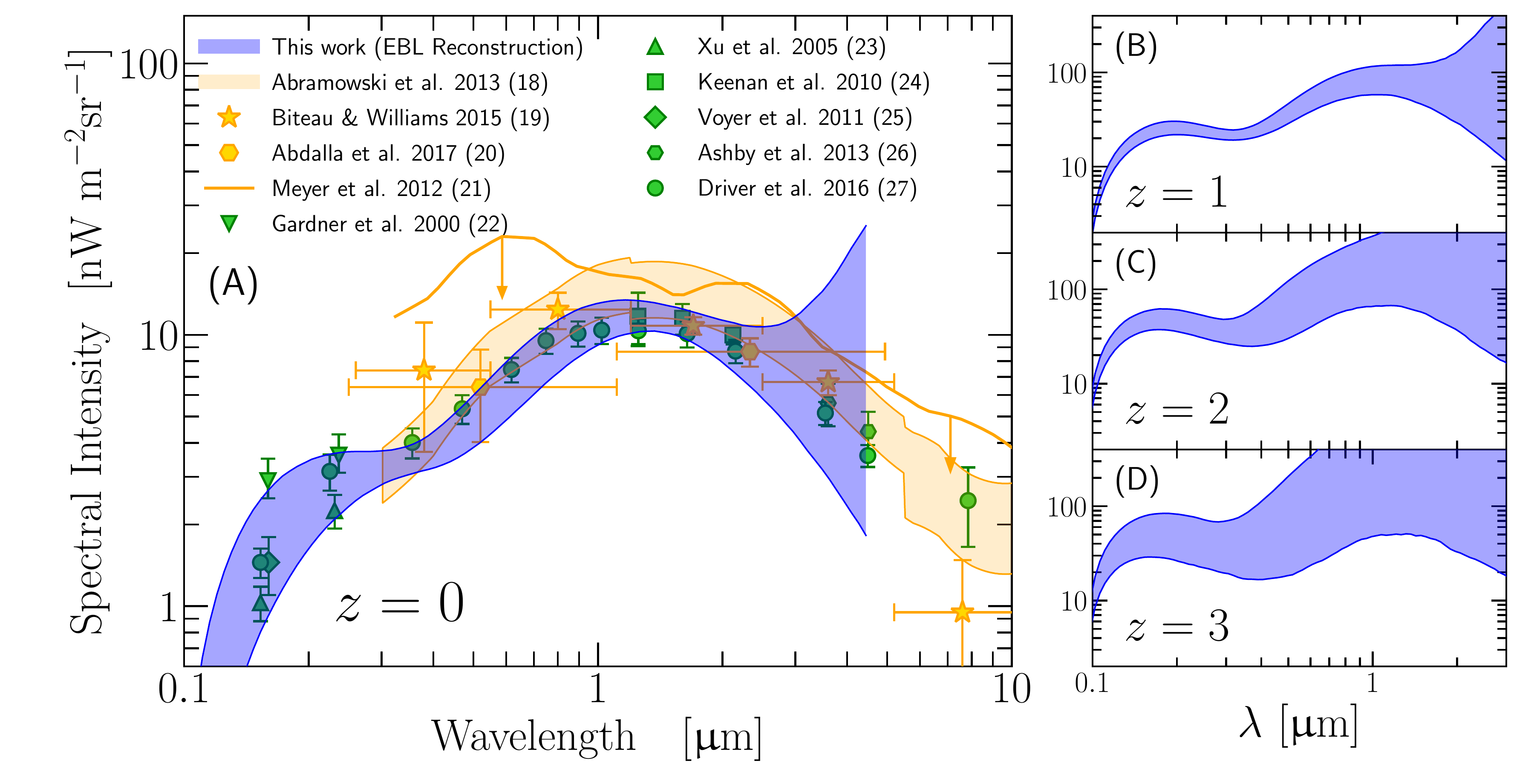}
\caption{{\bf The spectral intensity of the EBL in the Universe today (A) and
  at redshifts $z=1,2,3$ (B, C, and D)}. At $z=0$ data from other
  $\gamma$-ray based measurements are shown with orange symbols
\cite{abramowski13,biteau15,abdalla17,meyer12} while
  integrated galaxy counts are displayed with green symbols \cite{gardner00,xu05,keenan10,voyer11,ashby13, driver16}. 
 The blue areas show the 1\,$\sigma$
  confidence regions based on the reconstructed cosmic emissivity
  \cite{som}. At higher redshift (B, C, and D), the EBL is shown
    in physical coordinates. 
Figure \ref{fig:ebl_models} in \cite{som} includes a more complete set of
    measurements from the literature. 
\label{fig:ebl}}
\end{figure}

Our analysis leads to detections of the EBL
attenuation across the entire $0.03<z<3.1$ redshift range of the
blazars. From this, we identify
 the redshift at which, for a given energy, the Universe becomes
opaque to $\gamma$ rays, known as {cosmic $\gamma$-ray
horizon} (Figure~\ref{fig:horizon}). 
With the optical depths measured in six
energy bins ($10-1000$ GeV) across twelve redshift bins \cite{som} we are able to
reconstruct the intensity of the EBL at
different epochs (Figure~\ref{fig:ebl}).
We model the cosmic emissivity (luminosity density) of sources as several
simple spectral components at  UV,
optical, and near-IR (NIR) wavelengths. These components are  allowed
to vary in amplitude and
evolve with redshift independently of
each other to reproduce, through a Markov Chain Monte Carlo  (MCMC) analysis, the
optical depth data. The emissivities as a function of
wavelength and redshift allow us to reconstruct the history of the EBL
over $\sim90$\,\% of cosmic time.

At $z=0$ the energy spectrum of the EBL is close
to the one inferred by resolving individual galaxies 
in deep fields \cite{driver16}. At all other epochs, {\it
  Fermi} LAT is most sensitive to the UV-optical component
of the EBL, and is only able to constrain the NIR component at more recent times
(see Figure~\ref{fig:ebl}).
The intensity of the UV background in the local Universe remains
uncertain, with independent studies reporting differing values
\cite{gardner00,xu05,voyer11}.
Our determination of 2.56$^{+0.92(2.23)}_{-0.87(1.49)}$\,nW m$^{-2}$
sr$^{-1}$,   1$\sigma$(2$\sigma$),  at 0.2\,$\upmu$m favors an intermediate UV intensity in agreement with
\cite{voyer11}. 
In the NIR our measurement of $11.6^{+1.3(2.6)}_{-1.4(3.1)}~{\rm
  nW~m^{-2}sr^{-1}}$, 1$\sigma$ (2$\sigma$), at
  1.4\,$\upmu$m is consistent with integrated galaxy counts
\cite{keenan10,ashby13}, leaving little room for additional
components, contrary to some suggestions
\cite{bernstein07,matsuura17}. This notably includes contributions from stars
that have been stripped from galaxies as the technique presented here is
sensitive to all photons \cite{zemcov14,burke15}.

\begin{figure}
\centering
  \includegraphics[width=1\textwidth]{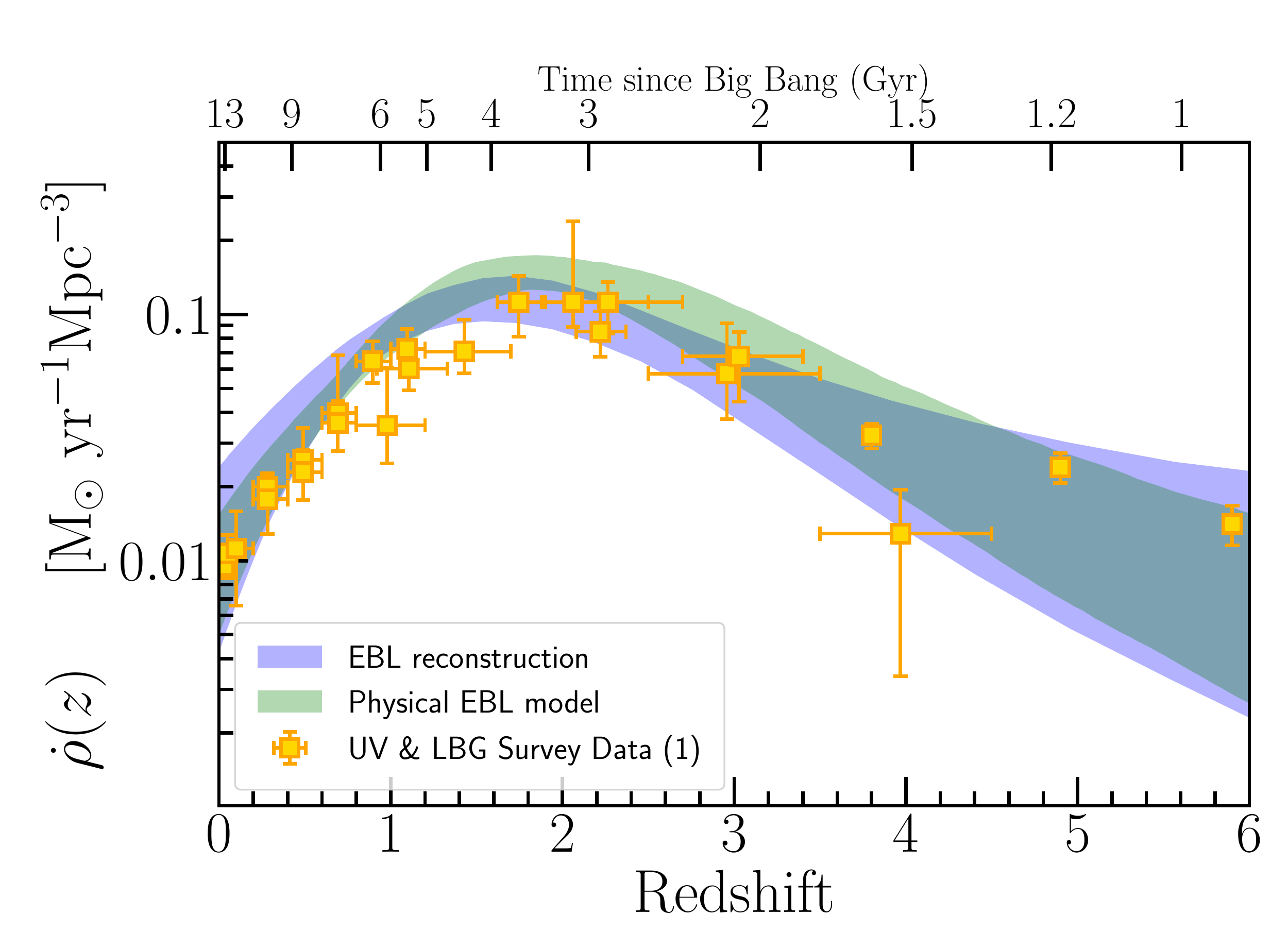}
\caption{{\bf The cosmic star-formation history as constrained from the optical depth data.} The
  shaded regions correspond to the 1$\sigma$ confidence regions on the
  star formation rate density as a function of redshift, $\dot{\rho}(z)$,
  obtained from two independent methods, based on 1) a physical EBL
  model (green) and 2) an empirical EBL
  reconstruction (blue, see \cite{som}). The data points show the
  SFH derived from UV surveys at low $z$ and deep Lyman Break Galaxy
  (LBG) surveys at high-$z$ (see  review of \cite{Madau2014} and
  references therein). Figure \ref{fig_sfh_all} in \cite{som} includes a more complete set of data from different tracers
  of the star-formation rate.
\label{fig:sfr}}
\end{figure}

At any epoch, the EBL is composed of the emission of all
stars \cite{note2}
that existed up to that
point in time and  can therefore be used to infer properties related to the evolution of galaxy populations.
We focus on the cosmic SFH, which we
determine using two independent methods. First, we use the
reconstructed UV emissivity across cosmic time to derive the SFH from
established relations between the UV luminosity and star-formation
rate \cite{kennicutt98}, taking into account the mean dust
extinction within galaxies
\cite{Bouwens2012,Burgarella13,Andrews17}. 
The second approach uses a physical EBL model \cite{finke10} to calculate the
optical depth due to the EBL directly from the SFH. The SFH is then
optimized using a MCMC to reproduce the {\it Fermi}-LAT optical depth
data \cite{som}.
The two approaches yield consistent results for the SFH, which is well
constrained out to a redshift of $z\approx 5$, i.e., to the epoch
1.5 billion years after the Big Bang (Figure~\ref{fig:sfr}).

Because the optical depth increases with the distance traveled by the $\gamma$ rays,
we obtain the tightest constraints in the redshift range $0.1<z<1.5$, beyond which
our sensitivity decreases due to the lower number of observed blazars. To improve the constraint of
the SFH beyond $z=3$, we have complemented the blazar sample with 
a gamma-ray burst (GRB~080916C) at $z=4.35$
\cite{lat_grb09}.
This allows us to place upper limits on the SFH at
  $z\gsim5$, because photons generated at redshifts higher than the
  $z=4.35$ limit of our sample remain in the EBL, become redshifted,  
  and start interacting with the $\gamma$ rays from the blazars
  and the GRB used here at $z<4.35$.

\begin{figure}
\centering
  \includegraphics[width=0.65\textwidth]{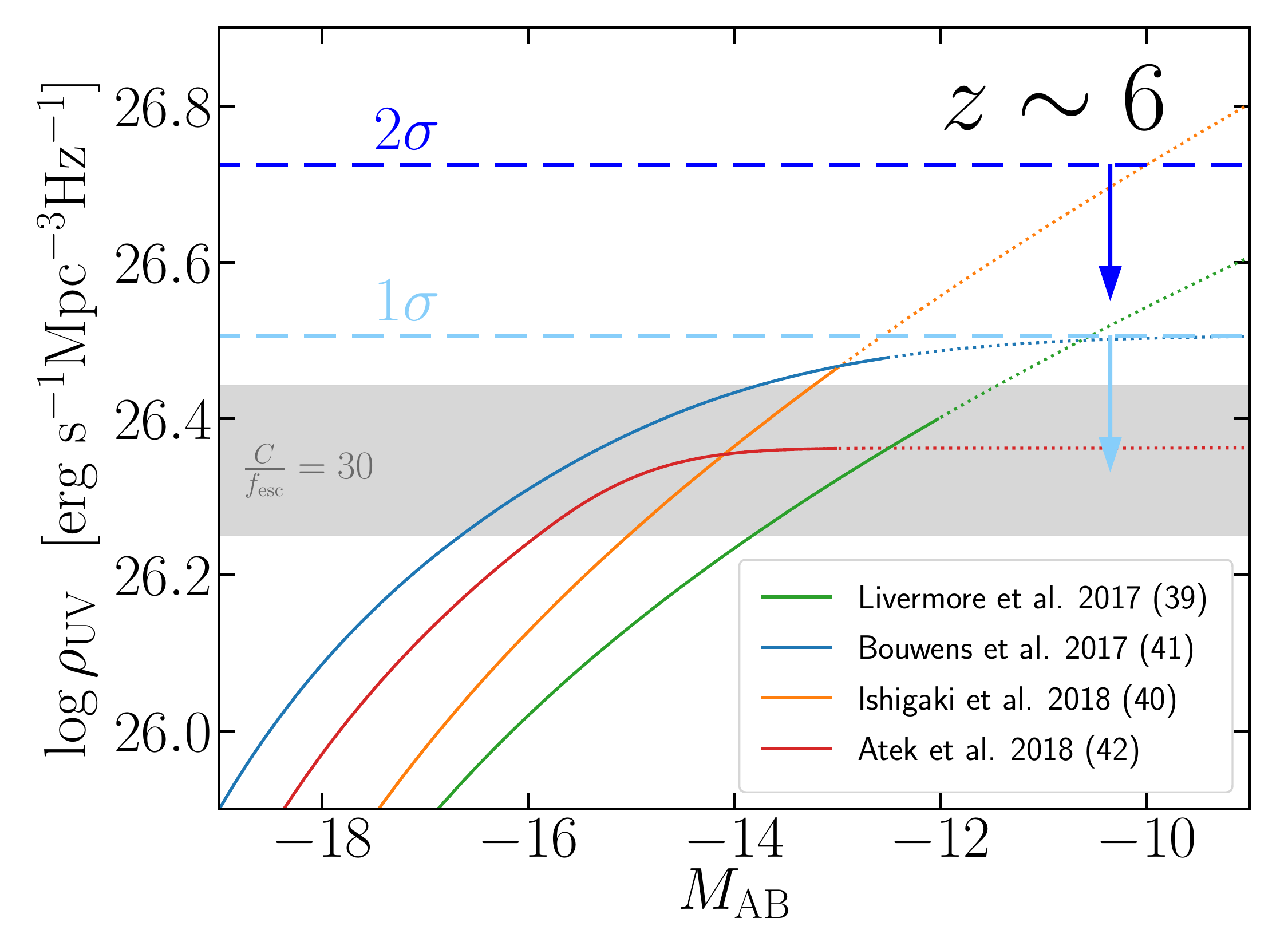}
\caption{ {\bf Upper limits on the UV luminosity density of
    galaxies at $z\sim
  6$}. The 1\,$\sigma$ and 2\,$\sigma$ limits
  are shown as dashed horizontal lines, light blue and dark blue
  respectively. The solid
  curves show the $z\sim 6$ UV emissivity from \cite{Livermore17,Bouwens17,Ishigaki18,Atek18} of the Hubble Frontier
  Fields (HFF) program as a function of the lower integration limit of the UV
  luminosity function. The dotted lines correspond to extrapolations
  beyond the limiting magnitude of the HFF analyses. The
  data from \cite{Bouwens17} correspond to their ``GLAFIC'' case. The lines of
  \cite{Ishigaki18} and \cite{Atek18} have been shifted up by 0.15 dex
  to account for evolution of their combined $z\sim 6-7$ sample to $z\sim 6$
  . The grey area
  corresponds to the luminosity required to keep the Universe ionized
  at $z=6$ assuming $C/f_{\rm esc}=30$, where $C$ is the clumping factor
  of ionized hydrogen and $f_{\rm esc}$ 
is the mean escape fraction of ionizing photons  \cite{som}.}
\label{fig_Mcut}
\end{figure}

At $z\gtrsim6$ the far-UV background (photon energy $>13.6$\,eV) is responsible for the re-ionization
of the neutral hydrogen in the Universe, but the nature of ionizing sources
has not been conclusively identified. One possibility is that ultra-faint galaxies existing in
large numbers can provide the required ionizing photons 
\cite{finkelstein2015,robertson2015}.
In this case, the galaxy
UV luminosity function must be steep at the faint end.
Recent measurements of the luminosity function in the
deepest {\it Hubble} fields remain inconclusive at the faintest levels
(absolute AB magnitude $M_{\rm
  AB}\gsim -15$) with some suggesting a continued steep faint-end
slope \cite{Livermore17,Ishigaki18} and others claiming a turnover \cite{Bouwens17,Atek18}. Our upper limits 
at $z=5-6$ on the UV emissivity  $\rho_{\rm UV} < 3.2(5.3) \times
10^{26}$ erg s$^{-1}$ Mpc$^{-3}$ Hz$^{-1}$
1\,$\sigma$ (2\,$\sigma$), see Figure~\ref{fig_Mcut},  suggest a turnover of the
luminosity function at $M_{\rm AB}\sim -14$ in agreement with
\cite{Bouwens17} and \cite{Atek18}. This still allows for abundant
photons to drive the re-ionization.

\bibliographystyle{Science}

\section*{Acknowledgments}

\subsection*{Funding}
M.~Ajello, V.~Paliya, and A.~Desai acknowledge support from NSF and NASA through grants
AST-1715256, NNX16AR72G, and 80NSSC17K0506.
K.~Helgason acknowledges support from the Icelandic Research Fund, grant
number 173728-051.
J.~Finke was supported by the Chief of Naval Research and by a grant of computer time from the Department of Defense High Performance Computing Modernization Program at the Naval Research Laboratory.
A.~Dom\'{i}nguez  thanks the support of the Juan de la Cierva program from the
Spanish MEC.

\noindent {\textbf{{\textit  Fermi}-LAT collaboration:}} The \textit{Fermi} LAT Collaboration acknowledges generous ongoing support
from a number of agencies and institutes that have supported both the
development and the operation of the LAT as well as scientific data analysis.
These include the National Aeronautics and Space Administration and the
Department of Energy in the United States, the Commissariat \`a l'Energie Atomique
and the Centre National de la Recherche Scientifique / Institut National de Physique
Nucl\'eaire et de Physique des Particules in France, the Agenzia Spaziale Italiana
and the Istituto Nazionale di Fisica Nucleare in Italy, the Ministry of Education,
Culture, Sports, Science and Technology (MEXT), High Energy Accelerator Research
Organization (KEK) and Japan Aerospace Exploration Agency (JAXA) in Japan, and
the K.~A.~Wallenberg Foundation, the Swedish Research Council and the
Swedish National Space Board in Sweden. 
Additional support for science analysis during the operations phase is gratefully
acknowledged from the Istituto Nazionale di Astrofisica in Italy and the Centre
National d'\'Etudes Spatiales in France. This work performed in part under DOE
Contract DE-AC02-76SF00515. {J. Conrad acknowledge funding from Wallenberg Academy.
M. Razzano is funded by contract FIRB-2012-RBFR12PM1F from the Italian Ministry of Education, University and Research (MIUR)}

\subsection*{Author contribution:} All authors meet the journal
authorship criteria. M.~Ajello designed the project and wrote most
of the paper. V.~Paliya performed the analysis of the $\gamma$-ray data, while
K.~Helgason derived the constrains on the EBL and the star-formation history and wrote all
the corresponding text. J.~Finke derived the results of the stellar
population method and wrote the corresponding text. A.~Desai tested all the EBL models while
A.~Dom\'{i}nguez  provided all the EBL-related data reported in the figures and
wrote the corresponding text.  All
co-authors have read, provided comments and approved the manuscript.

\subsection*{Competing interests:} All co-authors declare that there
are no competing interesting.

\subsection*{Data and materials availability:} The data used to derive
the results presented in this paper are provided in tabular form in
the supplementary materials.
The {\it Fermi}-LAT data and software needed to analyze those are available from the {\it Fermi} Science
Support Center \url{http://fermi.gsfc.nasa.gov/ssc}. 
The reconstructed optical depth templates, EBL, and SFH
are also available at \url{https://figshare.com/s/14f943002230d69a4afd}.
The tool to produce physical models of blazars' SEDs is available at \url{www.isdc.unige.ch/sedtool}.

\subsection*{Supplementary Materials}
\url{www.sciencemag.org}\\
Authors and Affiliations\\
Materials and Methods\\
Figs. S1 to S12\\
Table S1 to S5\\
References (43-115)\\

\clearpage

%
%

\renewcommand{\thefigure}{S\arabic{figure}}
\renewcommand{\thetable}{S\arabic{table}}
\renewcommand{\theequation}{S\arabic{equation}}

\setcounter{figure}{0}
\setcounter{table}{0}
\setcounter{equation}{0}

\section*{Authors and Affiliations}
\noindent
S.~Abdollahi$^{1}$, 
M.~Ackermann$^{2}$, 
M.~Ajello$^{3}\dagger$, 
W.~B.~Atwood$^{4}$, 
L.~Baldini$^{5}$, 
J.~Ballet$^{6}$, 
G.~Barbiellini$^{7,8}$, 
D.~Bastieri$^{9,10}$, 
J.~Becerra~Gonzalez$^{11,12}$, 
R.~Bellazzini$^{13}$
% ...
\medskip
\begin{enumerate}
\item[1.] Department of Physical Sciences, Hiroshima University, Higashi-Hiroshima, Hiroshima 739-8526, Japan
\item[2.] Deutsches Elektronen Synchrotron DESY, D-15738 Zeuthen, Germany
\item[3.] Department of Physics and Astronomy, Clemson University, Kinard Lab of Physics, Clemson, SC 29634-0978, USA
\item[4.] Santa Cruz Institute for Particle Physics, Department of Physics and Department of Astronomy and Astrophysics, University of California at Santa Cruz, Santa Cruz, CA 95064, USA
\item[5.] Universit\`a di Pisa and Istituto Nazionale di Fisica Nucleare, Sezione di Pisa I-56127 Pisa, Italy
\item[6.] Laboratoire Astrophysique Interactions Multi-\'echelles,  Commissariat \'a l'\'energie atomique-Institute of Research into the Fundamental Laws of the Universe/Centre national de la recherche scientifique/Universit\'e Paris Diderot, Service d'Astrophysique, Commissariat \'a l'\'energie atomique Saclay, F-91191 Gif sur Yvette, France
\item[7.] Istituto Nazionale di Fisica Nucleare, Sezione di Trieste, I-34127 Trieste, Italy
\item[8.] Dipartimento di Fisica, Universit\`a di Trieste, I-34127 Trieste, Italy
\item[9.] Istituto Nazionale di Fisica Nucleare, Sezione di Padova, I-35131 Padova, Italy
\item[10.] Dipartimento di Fisica e Astronomia ``G. Galilei'', Universit\`a di Padova, I-35131 Padova, Italy
\item[11.] National Aeronautics and Space Administration Goddard Space Flight Center, Greenbelt, MD 20771, USA
\item[12.] Department of Astronomy, University of Maryland, College Park, MD 20742, USA
\item[13.] Istituto Nazionale di Fisica Nucleare, Sezione di Pisa, I-56127 Pisa, Italy
\item[14.] Dipartimento di Fisica ``M. Merlin" dell'Universit\`a e del Politecnico di Bari, I-70126 Bari, Italy
\item[15.] Istituto Nazionale di Fisica Nucleare, Sezione di Bari, I-70126 Bari, Italy
\item[16.] W. W. Hansen Experimental Physics Laboratory, Kavli Institute for Particle Astrophysics and Cosmology, Department of Physics and SLAC National Accelerator Laboratory, Stanford University, Stanford, CA 94305, USA
\item[17.] Istituto Nazionale di Fisica Nucleare, Sezione di Torino, I-10125 Torino, Italy
\item[18.] Dipartimento di Fisica, Universit\`a degli Studi di Torino, I-10125 Torino, Italy
\item[19.] Department of Physics and Astronomy, University of Padova, Vicolo Osservatorio 3, I-35122 Padova, Italy
\item[20.] Laboratoire Univers et Particules de Montpellier, Universit\'e Montpellier, Centre national de la recherche scientifique/Institut national de physique nucl\'eaire de physique des particules, F-34095 Montpellier, France
\item[21.] Laboratoire Leprince-Ringuet, \'Ecole polytechnique, Centre national de la recherche scientifique/Institut national de physique nucl\'eaire de physique des particules, F-91128 Palaiseau, France
\item[22.] Center for Research and Exploration in Space Science and Technology (CRESST) and National Aeronautics and Space Administration Goddard Space Flight Center, Greenbelt, MD 20771, USA
\item[23.] Istituto Nazionale di Astrofisica-Istituto di Astrofisica Spaziale e Fisica Cosmica Milano, via E. Bassini 15, I-20133 Milano, Italy
\item[24.] Italian Space Agency, Via del Politecnico snc, 00133 Roma, Italy
\item[25.] Space Science Division, Naval Research Laboratory, Washington, DC 20375-5352, USA
\item[26.] Space Science Data Center - Agenzia Spaziale Italiana, Via del Politecnico, snc, I-00133, Roma, Italy
\item[27.] Istituto Nazionale di Fisica Nucleare, Sezione di Perugia, I-06123 Perugia, Italy
\item[28.] Department of Physics and Astronomy, Sonoma State University, Rohnert Park, CA 94928-3609, USA
\item[29.] Department of Physics, Stockholm University, AlbaNova, SE-106 91 Stockholm, Sweden
\item[30.] The Oskar Klein Centre for Cosmoparticle Physics, AlbaNova, SE-106 91 Stockholm, Sweden
\item[31.] Istituto Nazionale di Astrofisica Istituto di Radioastronomia, I-40129 Bologna, Italy
\item[32.] Dipartimento di Astronomia, Universit\`a di Bologna, I-40127 Bologna, Italy
\item[33.] Universit\`a Telematica Pegaso, Piazza Trieste e Trento, 48, I-80132 Napoli, Italy
\item[34.] Grupo de Altas Energ\'ias, Universidad Complutense de Madrid, E-28040 Madrid, Spain
\item[35.] Friedrich-Alexander-Universit\"at Erlangen-N\"urnberg, Erlangen Centre for Astroparticle Physics, Erwin-Rommel-Str. 1, 91058 Erlangen, Germany
\item[36.] Circolo Astrofili Talmassons, I-33030 Campoformido (UD), Italy
\item[37.] Laboratoire de Physique et Chimie de l'Environnement et de l'Espace -- Universit\'e d'Orl\'eans / Centre national de la recherche scientifique, F-45071 Orl\'eans Cedex 02, France
\item[38.] Station de radioastronomie de Nan\c{c}ay, Observatoire de Paris, Centre national de la recherche scientifique/Institut national des sciences de l'univers, F-18330 Nan\c{c}ay, France
\item[39.] The George Washington University, Department of Physics, 725 21st St, NW, Washington, DC 20052, USA
\item[40.] Max-Planck-Institut f\"ur Astrophysik, Postfach 1317, D-85741 Garching, Germany
\item[41.] Science Institute, University of Iceland, IS-107 Reykjavik, Iceland
\item[42.] Kungliga Tekniska h{\"o}gskolan Royal Institute of Technology and Stockholm University, Roslagstullsbacken 23, SE-106 91 Stockholm, Sweden
\item[43.] Department of Physics, Kungliga Tekniska h{\"o}gskolan Royal Institute of Technology, AlbaNova, SE-106 91 Stockholm, Sweden
\item[44.] Centre d'\'Etudes Nucl\'eaires de Bordeaux Gradignan, Institut national de physique nucl\'eaire de physique des particules/Centre national de la recherche scientifique, Universit\'e Bordeaux 1, BP120, F-33175 Gradignan Cedex, France
\item[45.] Hiroshima Astrophysical Science Center, Hiroshima University, Higashi-Hiroshima, Hiroshima 739-8526, Japan
\item[46.] Istituto Nazionale di Fisica Nucleare, Sezione di Roma ``Tor Vergata", I-00133 Roma, Italy
\item[47.] Department of Physics and Astronomy, University of Denver, Denver, CO 80208, USA
\item[48.] Max-Planck-Institut f\"ur Physik, D-80805 M\"unchen, Germany
\item[49.] Osservatorio Astronomico di Trieste, Istituto Nazionale di Astrofisica, I-34143 Trieste, Italy
\item[50.] Department of Physics, University of Johannesburg, PO Box 524, Auckland Park 2006, South Africa
\item[51.] Institut f\"ur Astro- und Teilchenphysik and Institut f\"ur Theoretische Physik, Leopold-Franzens-Universit\"at Innsbruck, A-6020 Innsbruck, Austria
\item[52.] Department of Physics, The University of Hong Kong, Pokfulam Road, Hong Kong, China
\item[53.] Laboratory for Space Research, The University of Hong Kong, Hong Kong, China
\item[54.] Nycb Real-Time Computing Inc., Lattingtown, NY 11560-1025, USA
\item[55.] Purdue University Northwest, Hammond, IN 46323, USA
\item[56.] Solar-Terrestrial Environment Laboratory, Nagoya University, Nagoya 464-8601, Japan
\item[57.] Centre national de la recherche scientifique, Institut de Recherche en Astrophysique et Plan\'etologie, F-31028 Toulouse cedex 4, France
\item[58.] Galaxies Astrophysique des Hautes \'Energies Cosmologie, Universit de Toulouse, Universit\'e Paul Sabatier-Observatoire midi-pyr\'en\'ees, Institut de Recherche en astrophysique et plan\'etologie, F-31400 Toulouse, France
\item[59.] Institute of Space Sciences (Consejo Superior de Investigaciones Científicas-Institut d'Estudis Espacials de Catalunya), Campus Universitat Aut\'onoma de Barcelona, Carrer de Magrans s/n, E-08193 Barcelona, Spain
\item[60.] Instituci\'o Catalana de Recerca i Estudis Avan\c{c}ats (ICREA), E-08010 Barcelona, Spain
\item[61.] Istituto Nazionale di Fisica Nucleare-Istituto di Astrofisica Spaziale e Fisica Cosmica Bologna, via P. Gobetti 101, I-40129 Bologna, Italy
\item[62.] Dipartimento di Fisica, Universit\`a degli Studi di Perugia, I-06123 Perugia, Italy
\item[63.] INTErnational Gamma-Ray Astrophysics Laboratory Science Data Centre, CH-1290 Versoix, Switzerland
\item[64.] California State University, Los Angeles, Department of Physics and Astronomy, Los Angeles, CA 90032, USA
\item[65.] Praxis Inc., Alexandria, VA 22303, resident at Naval Research Laboratory, Washington, DC 20375, USA
\item[66.] Istituto Nazionale di Fisica Nucleare, Sezione di Trieste, and Universit\`a di Trieste, I-34127 Trieste, Italy
\item[67.] Center for Astrophysics and Cosmology, University of Nova Gorica, Nova Gorica, Slovenia
\item[$\dagger$] majello@g.clemson.edu, helgason@hi.is,
  vpaliya@g.clemson.edu,\\
  justin.finke@nrl.navy.mil, abhishd@g.clemson.edu, alberto@gae.ucm.es
\end{enumerate}

\section*{Materials and Methods}

%
%
\section*{Sample Selection and Data Analysis}
\label{sec:sample}

\begin{figure*}[ht!]
  \begin{center}
  \begin{tabular}{c}
  	 \includegraphics[scale=0.83,clip=true,trim=0 0 0 0]{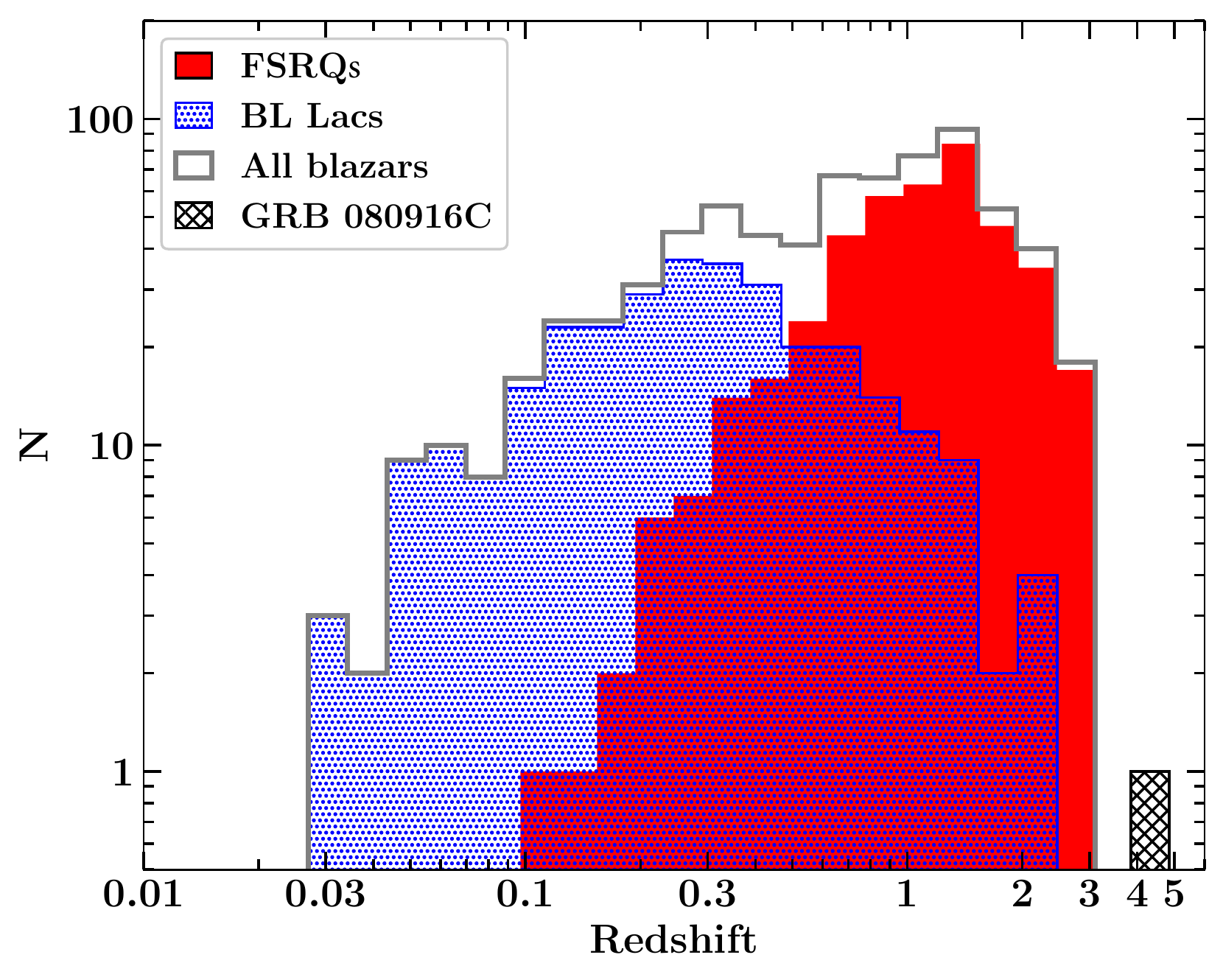} \\
\end{tabular}
  \end{center}
  \caption{{\bf Redshift distribution of the sources used in this analysis
    on a logarithmic scale.}
\label{fig:redshift}}
\end{figure*}

Our sample is selected starting from the objects reported in the third 
catalog of active galactic
nuclei detected by the LAT, 3LAC, \cite{3LAC}. We exclude all the
blazars reported there with a double association and those lacking a
redshift measurement. Most redshift
measurements for BL Lacs reported in 3LAC  come
from \cite{shaw13}. For each source we assess the significance of the
detection (between 1\,GeV and 1\,TeV) defining a test statistics ($TS$)
as $TS=2\Delta \log
\mathscr{L}$, where $\mathscr{L}$ represents the likelihood function
between models with and without the source of interest. We use this to
exclude all
sources that have a $TS<25$ in this analysis. Our final sample comprises {419}
FSRQs and {320} BL Lacs distributed (see Figure~\ref{fig:redshift}) between a redshift of 0.03 and 3.1.

The analysis relies on 101 months  (Aug. 2008 to Jan. 2017) of 
  Pass~8 (P8) class `SOURCE'
photons detected by the LAT between 1\,GeV and 1\,TeV. This dataset
was filtered to eliminate times when the spacecraft was over the
South Atlantic Anomaly and to remove photons detected at angles larger
than 100$^{\circ}$ from the zenith. For the analysis of each source
we use photons within 15$^{\circ}$ of the source position (region of interest, ROI). For each ROI we define a sky model that
comprises the diffuse Galactic \cite{diffuse_model} and extragalactic
emission \cite{templates} 
as well as the emission from background sources in the ROI. The latter
includes sources detected in the third {\it Fermi}-LAT catalog,  3FGL,
\cite{3FGL}
as well as any  new source that is detected because of the additional
exposure (with respect to the 3FGL) used here. These sources are found
generating a $TS$ map and identified as
excesses above a $TS=25$ threshold and added to the sky model with a
power-law spectrum. The LAT `P8R2$\_$SOURCE$\_$V6'
instrumental response function (IRF) and a binned likelihood method are used
to fit the sky model to the data. 

\subsection*{Intrinsic Spectra of Blazars}
\label{sec:int}
To capture the intrinsic curvature in the spectra of blazars we
adopt the following strategy that has been optimized  using
simulations prior to the data analysis (see below). The data are fitted only to a maximum energy up to which the
attenuation of the EBL is negligible. This is defined as the energy at
which the optical depth \tgg$<$0.1 for the model of
\cite{finke10}. However, we tested that our analysis is robust against
changes of EBL model used to define this maximum energy and changes to
the threshold (from e.g. \tgg$<0.1$ to \tgg$<0.05$). The optical
depth decreases sharply in this
regime where not many EBL photons are expected due to a characteristic drop-off at the
Lyman-limit (13.6\,eV).
Our baseline model for the intrinsic blazar spectrum is a log-parabola:
\begin{equation}
\frac{{\rm d}N}{{\rm d}E}=N_0 \left( \frac{E}{E_b}\right)^{-\alpha+\beta \log(E/E_b)}
\end{equation}
where $N_0$ (the normalization), $\alpha$ (photon index), $\beta$
(curvature) are all free parameters and $E_b$ is a scaling energy.
We also test whether an exponential power law could be a better
representation of the blazar spectrum
and this is defined as:
\begin{equation}
\frac{{\rm d}N}{{\rm d}E}=N_0 \left( \frac{E}{E_c} \right)^{\alpha} e^{-(E/E_b)^{\gamma_1}}
\end{equation}
where $E_c$ (cut
off energy) and $\gamma_1$ (the exponential index) are all free
parameters. Smoothed broken power laws and broken
  power laws were also tested, but they were never found to describe the blazar intrinsic
spectrum better than the two models reported above in the energy range
used in this work.

When testing  the exponential cut-off model we perform a first fit
with $\gamma_1$ fixed at 0.5 (justified from the observations of
hundreds of FSRQs, see \cite{ajello12_fsrq}) and then another fit
leaving $\gamma_1$ free to vary. We define two TS of curvature
$TS_{\rm c,1}$ and $TS_{\rm c,2}$ as follows:
\begin{eqnarray}
TS_{\rm c,1} & = & 2 (\log L_{{\rm exp},\gamma_1=0.5} - \log L_{\log-{\rm parabola}}) \\
TS_{\rm c,2} & = & 2 (\log L_{{\rm exp},\gamma_1={\rm free}} - \log L_{\log-{\rm parabola}}) .
\end{eqnarray}

where $\log L_{{\rm exp},\gamma_1=0.5}$ and $\log L_{{\rm exp},\gamma_1={\rm free}}$ are
the log-likelihoods derived using the exponential cut-off model with
$\gamma_1=0.5$ and $\gamma_1$ free to vary respectively and $\log
L_{\log-{\rm parabola}}$ is the log-likelihood of the log-parabola model.

We adopt the criteria reported in Table~\ref{tab:tsc} to choose the
model used to describe each blazar's intrinsic spectrum.
In order to avoid convergence problems, in the analysis presented above, the exponential index
$\gamma_1$ remains fixed at either 0.5 or the best-fitting value found at
this step. The median of the distribution of fitted $\gamma_1$
  values is  $\approx$0.5.

\begin{deluxetable}{ccc}
\tablewidth{0pt}
\tablecaption{Criteria, optimized on simulations, adopted to choose a spectral model.
\label{tab:tsc}}
\tablehead{\colhead{TS$_{c,1}$} & \colhead{TS$_{c,2}$} &
  \colhead{Model Chosen} 
}
\startdata
$<1$ & $<3$ & Log-parabola\\ 
$>1$ & $<3$ & Power law with exponential cut-off with $\gamma_1$=0.5\\
 & $>3$ & Power law with exponential cut-off with $\gamma_1$ free\\ 
\enddata
\end{deluxetable}

%
%
\section*{Analysis}
\label{sec:analysis}

\subsection*{Results for Blazars}
\label{sec:blazars}
Once the choice of the intrinsic spectrum for the sources has been
made, the analysis reverts to using the full, 1\,GeV--1\,TeV, energy
band and the modeled spectra of all sources include the EBL
attenuation as reported in Equation~\ref{eq:obs}, where $b$ is a parameter, common to all sources, that is
varied to fit the EBL model prediction to the data. A $b=1$ would mean
that the EBL model  predictions are in agreement with the LAT data,
while a $b=0$ would imply that there is no evidence for
attenuation due to  absorption by EBL photons in the spectra of
blazars.

\begin{figure*}[ht!]
  \begin{center}
  \begin{tabular}{c}
  	 \includegraphics[scale=0.83,clip=true,trim=0 0 0 0]{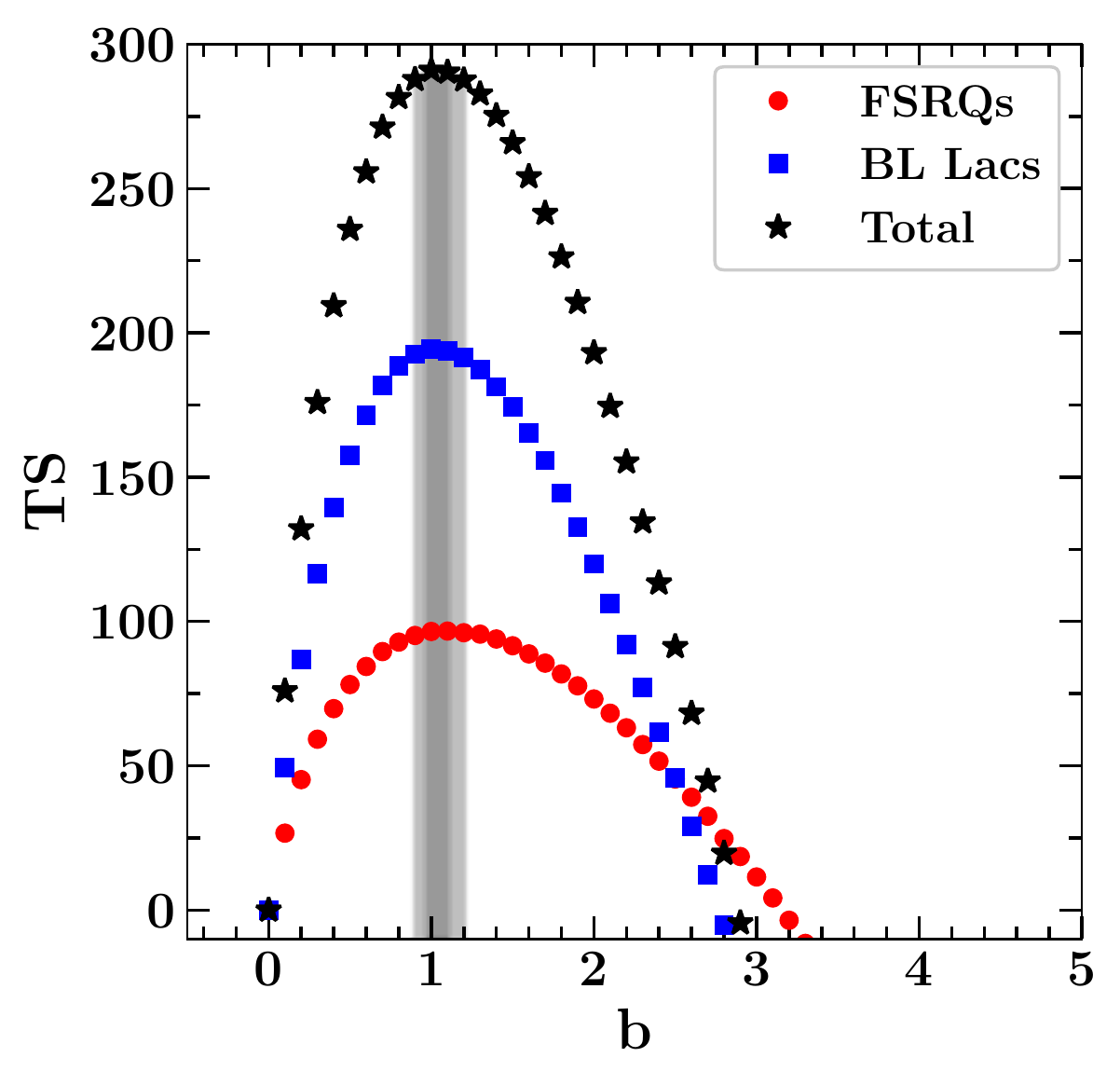} \\
\end{tabular}
  \end{center}
  \caption{{\bf Detection of the attenuation of the EBL.} Test statistics of the EBL as a function of the scaling
    parameter $b$ adopting the model of \cite{finke10}. The shaded regions show the 1\,$\sigma$
and 2\,$\sigma$ confidence intervals around the best fitting value of $b$. This $TS$
  profile was obtained by summing the $TS$ profiles of every source,
  including variable sources.
\label{fig:ts_vs_b}}
\end{figure*}

Because of the complexity of the problem, the $b$ parameter is not
optimized in one stage. Instead, for each source we scan the likelihood
function in very small steps of $b$ creating a
profile likelihood. In this process, the parameters of the diffuse
  emission, those of the brightest sources,  and those of the source of interest (except $\gamma_1$) are
  all left free to vary.
For each source,
the best-fitting $b$ value is the one that maximizes the
log-likelihood. A $TS$ of the detection of the EBL can be built
comparing the log-likelihood values at the best-fitting $b=b_0$ and at $b=0$
as $TS_{\rm EBL}=2[\log L(b_0) -\log L(b=0)]$. Because log-likelihoods (and
thus $TS$) are additive, we can determine the $b$ value that maximizes
the global (for all sources) likelihood and produces the largest
$TS_{\rm EBL}$.  In Figure~\ref{fig:ts_vs_b}, we plot the $TS$ profile, as a function of $b$, for all sources (and separately for BL Lacs and FSRQs) for the model of \cite{finke10}.
A $b$=1.03 improves the fit
by a $TS$ of $\sim$300, which corresponds to $\sim$17\,$\sigma$ for one
  degree of freedom. We note that the spectral
    evolution of the blazar class with redshift has a negligible
    effect on this analysis, as apparent from Figure~\ref{fig:ts_vs_b},
    which shows that the level of EBL measured using (mostly)
    hard-spectrum BL Lacs is in very good agreement with that found using 
soft-spectrum FSRQs. As an additional test we report the values of the
$b$ parameter for the model of \cite{finke10} for 
BL Lacs with a synchrotron peak frequency
  $>10^{16}$ Hz (called HSPs) and for the remaining BL Lacs; these are
respectively $b_{\rm HSPs}=0.98^{+0.09}_{-0.13}$ ($TS_{\rm EBL}$=125.8) and
$b_{\rm rest}=0.86^{+0.16}_{-0.10}$ ($TS_{\rm EBL}=45.1$). These highlight
once more that there is no bias in the level of the EBL due to the
spectral evolution of the blazar class.

One can also measure the compatibility of a model prediction with the
{\it Fermi}-LAT data defining a $TS$ as $TS_{b=1}=2[\log L(b_0)
-\log L(b=1)]$. By definition a large $TS_{b=1}$ implies that the model
predictions are in tension with the {\it Fermi}-LAT data; this
typically happens when the model predicts a larger-than-observed
attenuation. Table~\ref{tab:results} shows the results of our analysis
for some of the models available in the literature that have not been
found in tension with previous $\gamma$-ray data. The table shows that
the high model of \cite{scully2014} and the best-fitting model
of \cite{kneiske04} are excluded. Moreover,  the models of
\cite{gilmore12} and \cite{dominguez11}
are found in tension at the $\sim3$\,$\sigma$ level with the {\it
  Fermi}-LAT observations. All these models predict a larger optical-UV
intensity of the EBL than the models compatible with the LAT data.

\begin{deluxetable}{lccc||c}
\tablewidth{0pt}
\rotate
\tablecaption{Joint-likelihood results for different EBL models (first
  and second columns)
  ordered by decreasing value of the last column.  {The third column refers to the significance, in units of $\sigma$, of the 
attenuation in the spectra of blazars when a given
EBL model is scaled by the factor $b$. In this
case $b$=0 (i.e., no EBL absorption) constitutes the null
hypothesis. The fourth column lists the maximum likelihood values 
and 1\,$\sigma$ confidence ranges for the opacity scaling factor. In the last column, the $b$=1 case (i.e., EBL absorption as predicted 
by a given EBL model) constitutes the null hypothesis. This column shows
the compatibility (expressed in units of $\sigma$)
of the  predictions of EBL models with the {\it Fermi}-LAT
observations. Large values mean less likely to be compatible.}
\label{tab:results}}
\tablehead{\colhead{Model} & \colhead{Ref.} &
\colhead{Significance of $b$=0} &
\colhead{$b$ }          & 
\colhead{Significance of $b$=1} \\
\colhead{} & \colhead{} & \colhead{Rejection} & \colhead{} & \colhead{Rejection} 
}
\startdata
{\it Scully et al. (2014) -- high}         &  \cite{scully2014}           & 16.0 & 0.42$\pm0.03$ & 17.4 \\
{\it Kneiske et al. (2004) -- best -fit } &\cite{kneiske04}    & 16.9& 0.68$\pm0.05$ & 6.0   \\ 
{\it Gilmore et al. (2012) -- fixed}     &\cite{gilmore12}       &16.7 & 1.30$\pm0.10$ & 3.0   \\ 
{\it Gilmore et al. (2012) -- fiducial}  &\cite{gilmore12}     & 16.6& 0.81$\pm0.06$ & 2.9   \\ 
{\it Dominguez et al. (2011)}           &\cite{dominguez11} & 16.6 &
1.31$\pm0.10$ & 2.9   \\ 
{\it Franceschini et al. (2017)}           &\cite{franceschini17} & 16.4 &
1.25$\pm0.10$ & 2.5   \\ 
{\it Gilmore et al. (2009) }   &\cite{gilmore09}      & 16.7 & 1.03$\pm0.08$ & 2.4   \\ 
{\it Inoue et al. (2013)}              & \cite{inoue2013}     & 16.2 & 0.87$\pm0.06$ & 2.1   \\
{\it Kneiske \& Dole (2010)}            &\cite{kneiske10}   & 16.8 &0.94$\pm0.08$ & 1.7   \\  
{\it Helgason et al. (2012)}        &\cite{helgason12}                     & 16.5 & 1.10$\pm0.08$ & 1.3   \\ 
{\it Finke et al. (2010) --  model C}   &\cite{finke10}     & 17.1 & 1.03$\pm0.08$ & 0.4   \\ 
{\it Scully et al. (2014) -- low}          & \cite{scully2014}      & 16.0 & 1.00$\pm0.07$ & 0.1   \\

\enddata
\end{deluxetable}

\begin{figure*}[h]
  \begin{center}
  \begin{tabular}{c}
  	 \includegraphics[scale=0.73,clip=true,trim=0 0 0 0]{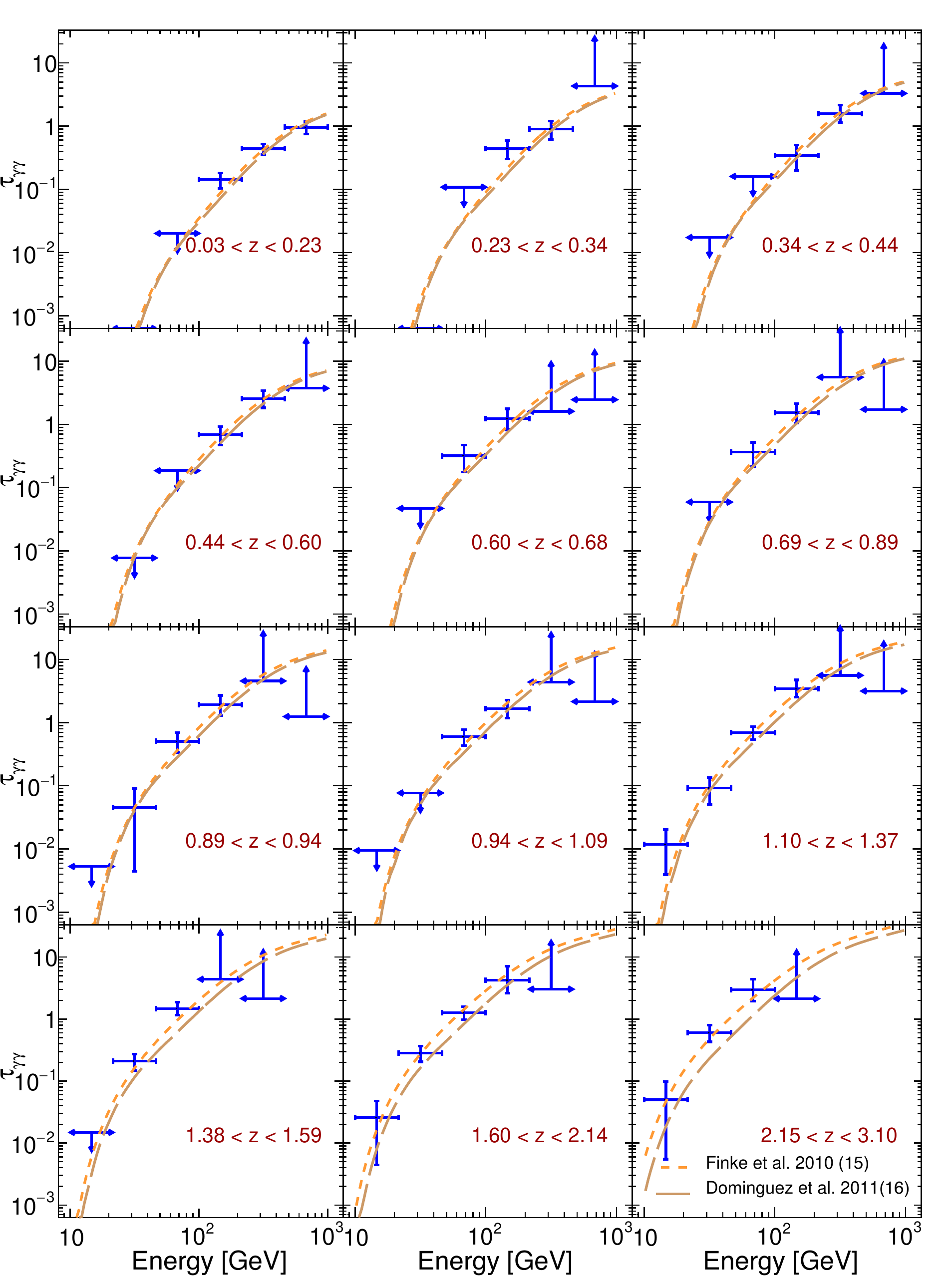} \\
\end{tabular}
  \end{center}
  \caption{{\bf Measurements of the optical depth $\tau_{\gamma\gamma}$ due
    to the EBL in different redshift and energy bins.} The lines show
  the predictions of two EBL models  \cite{finke10,dominguez11}.
\label{fig:taus}}
\end{figure*}

The optical depth as a function of energy and redshift can be measured
by repeating the above procedure  (i.e., renormalizing the optical depth predicted by
a model), but in small energy and redshift bins. In this process, the
  uncertainty due to the small disagreement between different EBL
  models, about the shape of the optical depth curve
within any given bin, has been included in the final uncertainty
of the optical depth. The final uncertainty includes also the 10\,\%
systematic uncertainty discussed below. The
redshift bins were chosen so that similar values of $TS_{\rm EBL}$ were
obtained in all the bins.
Figure~\ref{fig:taus} shows  measurements of the optical depth
$\tau_{\gamma\gamma}$ due to  EBL absorption in different
redshift and energy bins. 
It is  apparent from the figure that most of the
constraining power is around $\tau_{\gamma\gamma}\approx 1$. Formally the $\tau_{\gamma\gamma}(E,z)=1$ value marks the
cosmic $\gamma$-ray  horizon, i.e., the energy above which our
  Universe becomes opaque to $\gamma$ rays for a given redshift
  \cite{fazio70,dominguez13a}.
The energy at which $\tau_{\gamma\gamma}(E,z)=1$ at any redshift can be found by
renormalizing any EBL model to fit the data presented in
Figure~\ref{fig:taus} and propagating the (statistical plus systematic) uncertainties.
Figure~\ref{fig:horizon} shows that {\it Fermi} LAT maps the horizon
position with energy from low
($z\approx0.03$) to high ($z\approx3.1$)
redshift. Figure~\ref{fig:horizon} also shows the highest-energy
photons detected from the blazars in our sample.

\clearpage

\begin{deluxetable}{ccc||cccccc}
\tablewidth{0pt}
\rotate
\tablecaption{EBL Optical Depths $\tau_{\gamma\gamma}$ in bins of
  redshift and energy as reported in Figure~\ref{fig:taus}.
\label{tab:taus}}
\tablehead{\colhead{$\bar{z}$} & \colhead{$z_{min}$} &
  \colhead{$z_{max}$} & \colhead{[10.0-21.4]} &
  \colhead{[21.4-46.4]} & \colhead{[46.4-100.0]} &
  \colhead{[100.0-215.4]} & \colhead{[215.4-464.1]} &
  \colhead{[464.1-1000]} \\
  & & & \colhead{(GeV)}& \colhead{(GeV)}& \colhead{(GeV)}& \colhead{(GeV)}& \colhead{(GeV)}& \colhead{(GeV)}
}
\startdata
0.14 & 0.03 & 0.23  & \nodata & \nodata& $<0.02$& $0.14_{-0.04}^{+0.04}$& $0.44_{-0.09}^{+0.09}$& $0.96_{-0.21}^{+0.23}$\\ 
0.27 & 0.23 & 0.34  & \nodata & \nodata& $<0.11$& $0.44_{-0.14}^{+0.15}$& $0.90_{-0.28}^{+0.31}$& $>4.32$\\ 
0.39 & 0.34 & 0.44  & \nodata& $<0.02$& $<0.16$& $0.34_{-0.14}^{+0.16}$& $1.59_{-0.44}^{+0.56}$& $>3.30$\\ 
0.52 & 0.44 & 0.60  & \nodata& $<0.01$& $<0.19$& $0.69_{-0.22}^{+0.24}$& $2.55_{-0.73}^{+0.88}$& $>3.72$\\ 
0.65 & 0.60 & 0.68  & \nodata& $<0.05$& $0.32_{-0.14}^{+0.16}$& $1.24_{-0.42}^{+0.52}$& $>1.61$& $>2.46$\\ 
0.79 & 0.69 & 0.89 & \nodata& $<0.06$& $0.36_{-0.15}^{+0.16}$& $1.54_{-0.49}^{+0.61}$& $>5.57$& $>1.73$\\ 
0.92 & 0.89 & 0.94 & $<0.01$& $0.05_{-0.04}^{+0.04}$& $0.51_{-0.17}^{+0.18}$& $1.92_{-0.63}^{+0.78}$& $>4.56$& $>1.25$\\ 
1.01 & 0.94 & 1.09 & $<0.01$& $<0.08$& $0.60_{-0.16}^{+0.18}$& $1.67_{-0.50}^{+0.60}$& $>4.38$& $>2.16$\\ 
1.24 & 1.10 & 1.37 & $0.01_{-0.01}^{+0.01}$& $0.09_{-0.04}^{+0.04}$& $0.69_{-0.16}^{+0.17}$& $3.43_{-0.91}^{+1.32}$& $>5.59$& $>3.16$\\ 
1.47 & 1.38 & 1.59 & $<0.01$& $0.21_{-0.06}^{+0.06}$& $1.48_{-0.33}^{+0.38}$& $>4.38$& $>2.15$ & \nodata\\ 
1.82 & 1.60 & 2.14 & $0.03_{-0.02}^{+0.02}$& $0.28_{-0.08}^{+0.08}$& $1.27_{-0.30}^{+0.32}$& $4.21_{-1.59}^{+2.90}$& $>3.03$ & \nodata\\ 
2.40 & 2.15 & 3.10 & $0.05_{-0.04}^{+0.05}$& $0.60_{-0.17}^{+0.19}$& $2.97_{-1.02}^{+1.45}$& $>2.13$ & \nodata & \nodata\\ 
\enddata

\end{deluxetable}

\subsection*{GRB~080916C}
\label{sec:grb}
In order to constrain the EBL and SFH to the highest possible redshifts, we
complement the blazar sample with a single gamma-ray burst (GRB), GRB~080916C, detected by
{\it Fermi} LAT at $z$=4.35 \cite{lat_grb09,greiner09}. This was an
extremely luminous event, whose hard spectrum has already produced constraints on the SFH at high redshift
\cite{gilmore2012,inoue14}.  With respect to previous works, the
release of the {Pass~8} event-level analysis has allowed us to
recover more high-energy photons, particularly one at 27.4\,GeV,
$\approx$146\,GeV in the source frame, from
GRB~080916C \cite{atwood13}. 

The analysis is similar to the one reported by 
\cite{desai2017}. Transient-class photons
between 0.1\,GeV and 100\,GeV were downloaded  around a 10$^{\circ}$ position from the burst and from the
time of the GRB until 1775.9\,s later. Photons detected at a zenith
angle greater than 105$^{\circ}$  were removed.
The ROI model consists of the burst, the Galactic
and isotropic templates \cite{grb}.
We rely on the `P8R2\_TRANSIENT020' IRF.

\begin{figure*}[ht!]
  \begin{center}
  \begin{tabular}{c}
  	 \includegraphics[scale=0.8,clip=true,trim=0 0 0 0]{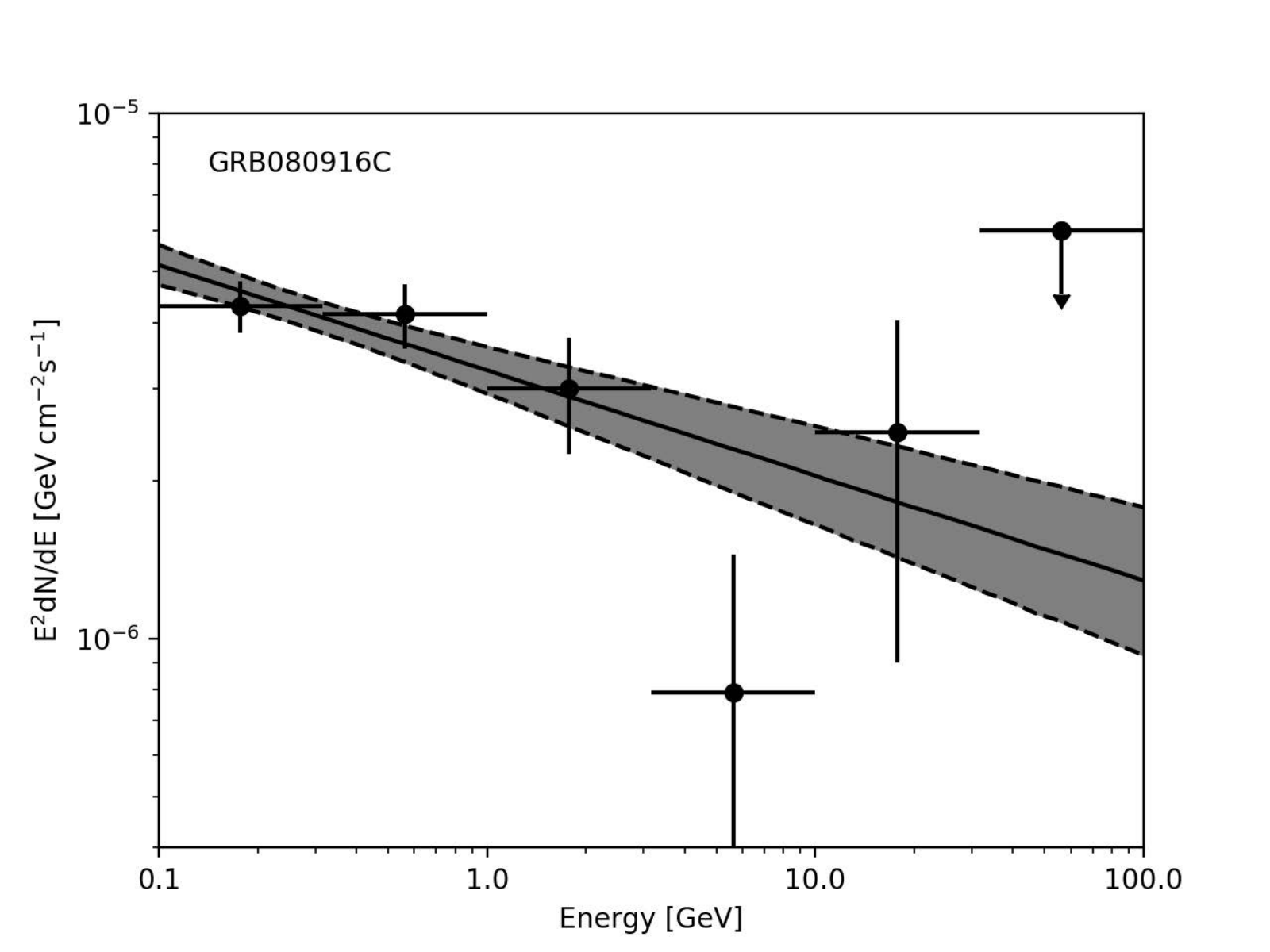} \\
\end{tabular}
  \end{center}
  \caption{{\bf Spectrum of GRB~080916C between 0.1\,GeV and
      100\,GeV.} The solid line and shaded region represent the
    best-fitting power-law model and its 1\,$\sigma$ uncertainty, respectively.
\label{fig:grb080916c}}
\end{figure*}

The source intrinsic spectrum is represented (and fitted up to 10\,GeV) by a single power law
(with a photon index of 2.25$\pm0.06$)
employing a time-averaged analysis. No curvature is
observed in the {\it Fermi}-LAT spectrum. A  time-resolved analysis
does not yield any difference for this work \cite{desai2017}. We produce a 95\,\% upper limit on the optical depth
by adopting the same method as described above for blazars. This
upper limit is $\tau_{\gamma\gamma}<0.46$ at an energy of
$\approx$17\,GeV and $z$=4.35 and it does not depend on the EBL model used
to  derive it. This upper limit is a factor of two lower than that
used by \cite{inoue14}. This is due to the additional photons detected
at $>$10\,GeV and particularly to the 27.4\,GeV photon.
The
  probability that this photon belongs to the background, rather than
  to GRB~080916C, is only 5$\times10^{-5}$. A so-called
`maximally conservative upper limit' based on the assumption that the
intrinsic spectrum cannot be harder than a power law with an index
$\Gamma=1.5$ is even more constraining; however, it is not
adopted here. 
The spectrum of GRB~080916C used here is shown in
Figure~\ref{fig:grb080916c}. 

%
%
\section*{Tests and Simulations}
\label{sec:tests}

\subsection*{Simulations of Blazar SEDs}
\label{sec:simulations}
The analysis chain described in the previous section has  first
  been tested and optimized on Monte Carlo simulation of synthetic
spectral-energy distributions (SEDs) of blazars with properties matching those of blazars observed by
{\it Fermi} LAT.

The SEDs are generated from physical models of blazars' emission that
include synchrotron and synchrotron self-Compton
as well as (for FSRQs)
external Compton scattering and were generated with the numerical code
presented in \cite{tramacere2009} and \cite{tramacere11}. These SEDs
reproduce the range of peak frequencies very well (for both the
synchrotron and $\gamma$-ray components), including peak curvatures and
$\gamma$-ray photon indices observed in both {\it Fermi}-LAT BL Lacs
and FSRQs. They include all known effects that contribute to 
  determining the curvature of the intrinsic $\gamma$-ray spectrum of
{\it Fermi} blazars. The crucial transition from the Thomson to the
Klein-Nishina cross section as well as (mostly important for FSRQs)
absorption within the broad line region (for different distances of
the emission region from the broad line region) are all taken into
account and contribute to determine the shape of the blazars' spectra
at high energy.

\begin{figure*}[ht!] 
  \begin{center}
  \begin{tabular}{c}
  	 \includegraphics[scale=0.83,clip=true,trim=0 0 0 0]{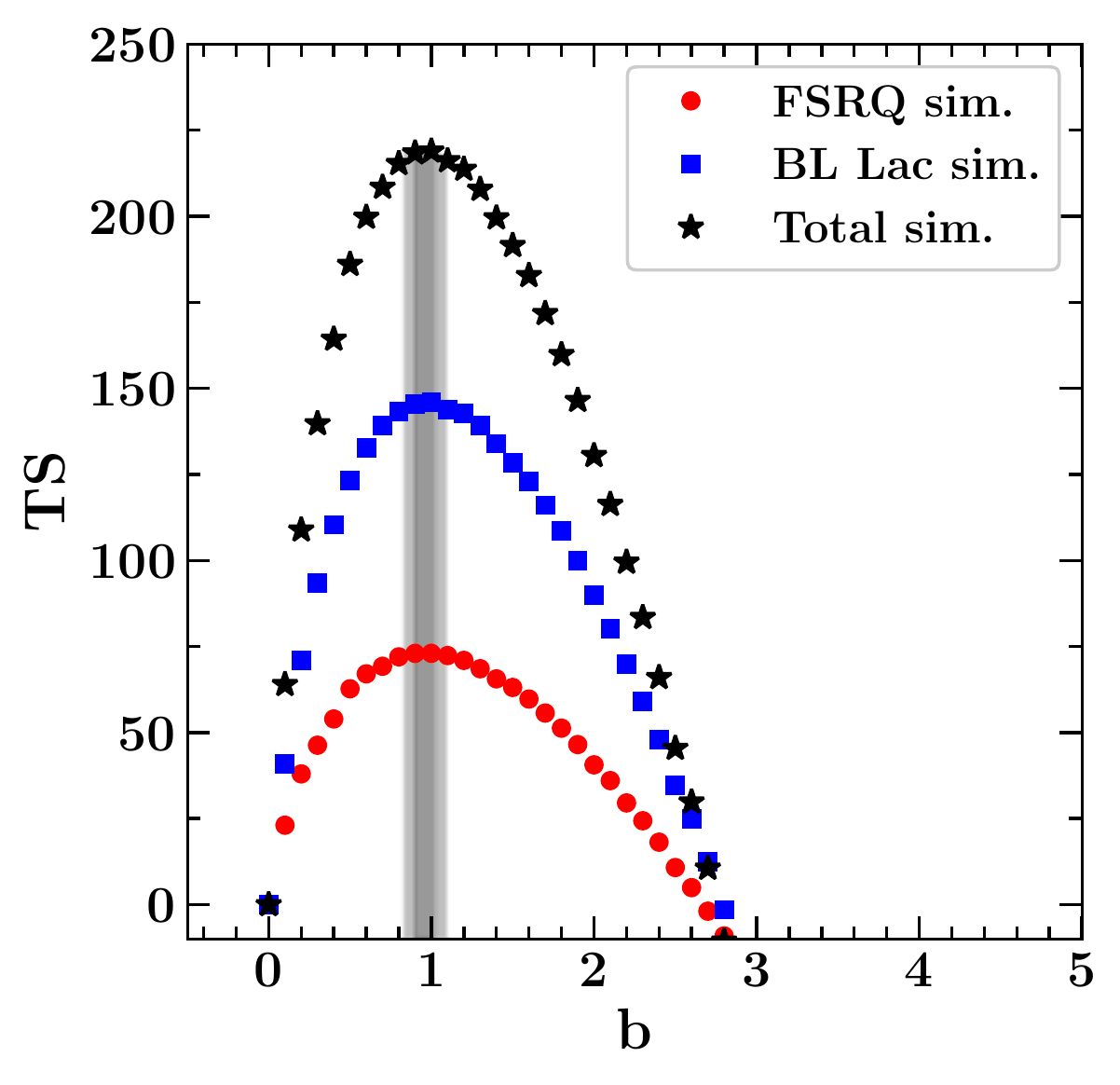} \\
\end{tabular}
  \end{center}
  \caption{{\bf Detection of the EBL attenuation in Monte Carlo
      simulations.}
Test statistics of the EBL as a function of the scaling
    parameter $b$ adopting the model of \cite{finke10} for our set of
    Monte Carlo simulations. The shaded regions show the 1\,$\sigma$
and 2\,$\sigma$ confidence intervals around the best fit.
\label{fig:simulations}}
\end{figure*}

These SEDs, attenuated by the EBL for a range of redshifts similar to
those of Figure~\ref{fig:redshift},  are then used to simulate LAT observations of these
synthetic sources and have been used to optimize the analysis set-up
presented above. In particular, the values of the
minimum energy ($E_{\rm min}$=1\,GeV) and those of $TS_{\rm c,1}$ and
$TS_{\rm c,2}$ have been derived from the analysis of simulations.
Figure~\ref{fig:simulations} shows
that the analysis chain employed in this work can effectively recover
the simulated level of EBL.

\subsection*{Variability}
\label{sec:variability}

\begin{figure*}[ht!]
  \begin{center}
  \begin{tabular}{cc}
  	 \includegraphics[scale=0.55,clip=true,trim=0 0 0
    0]{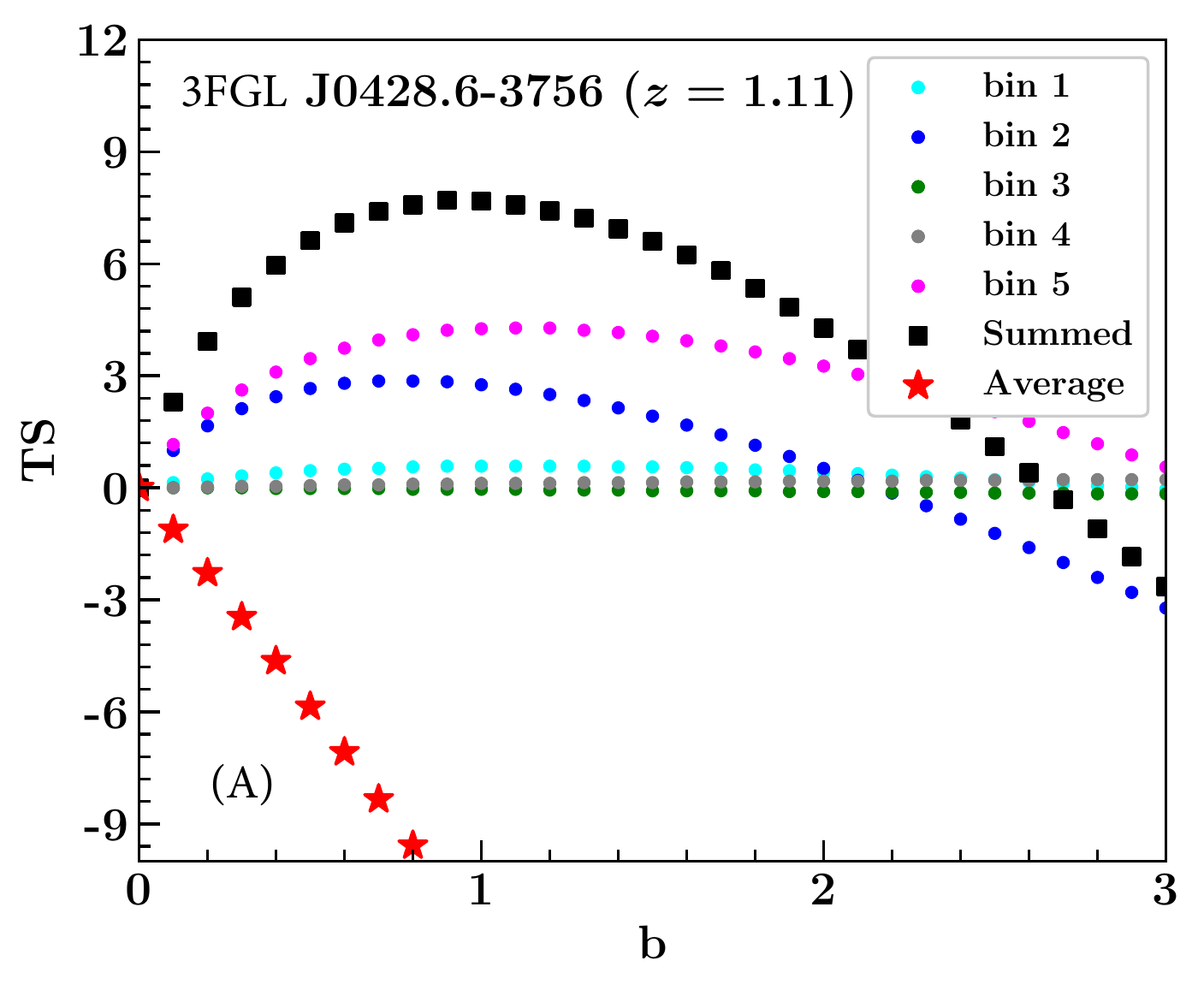} &
  	 \includegraphics[scale=0.55,clip=true,trim=0 0 0
    0]{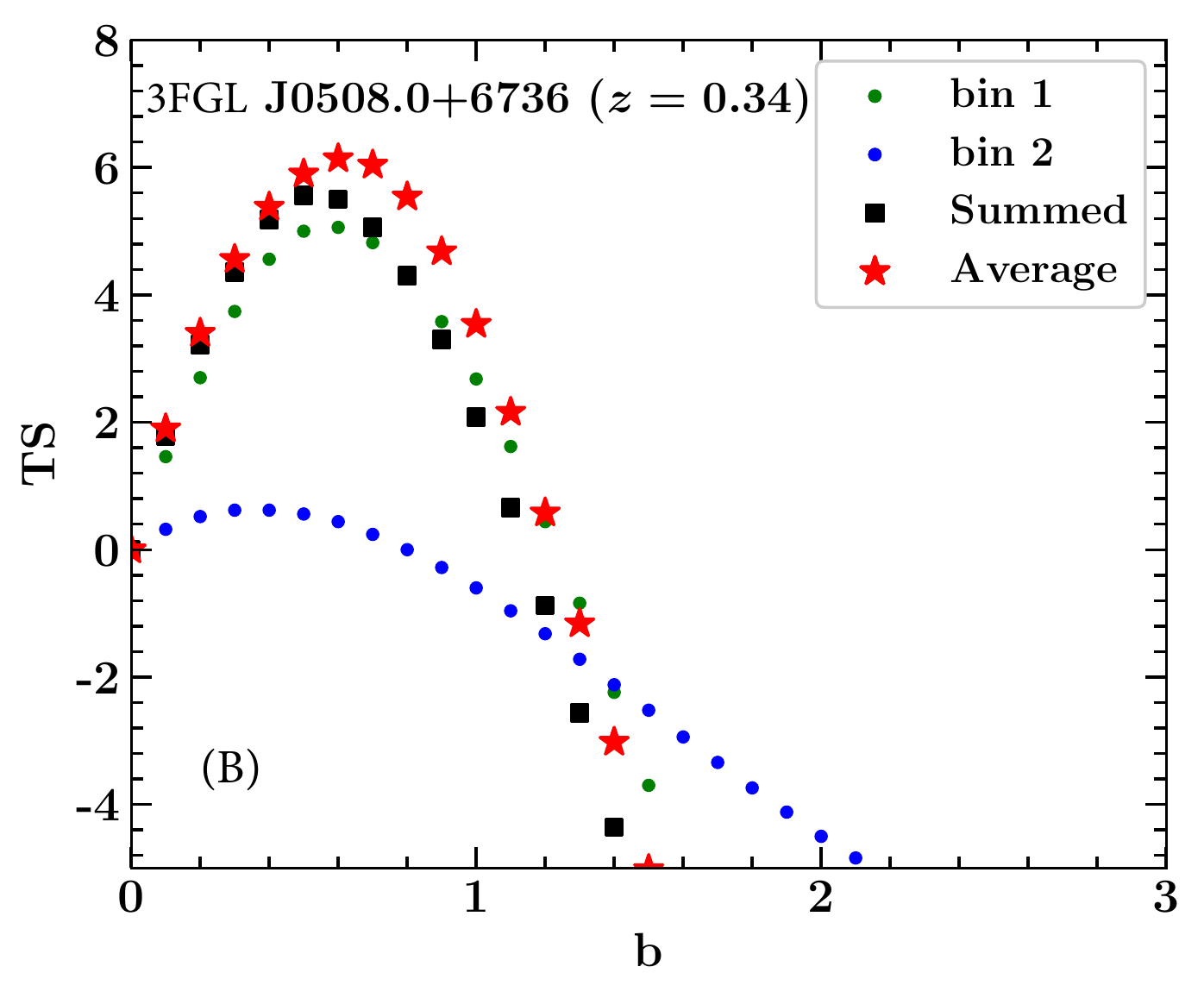} \\
  	 \includegraphics[scale=0.55,clip=true,trim=0 0 0
    0]{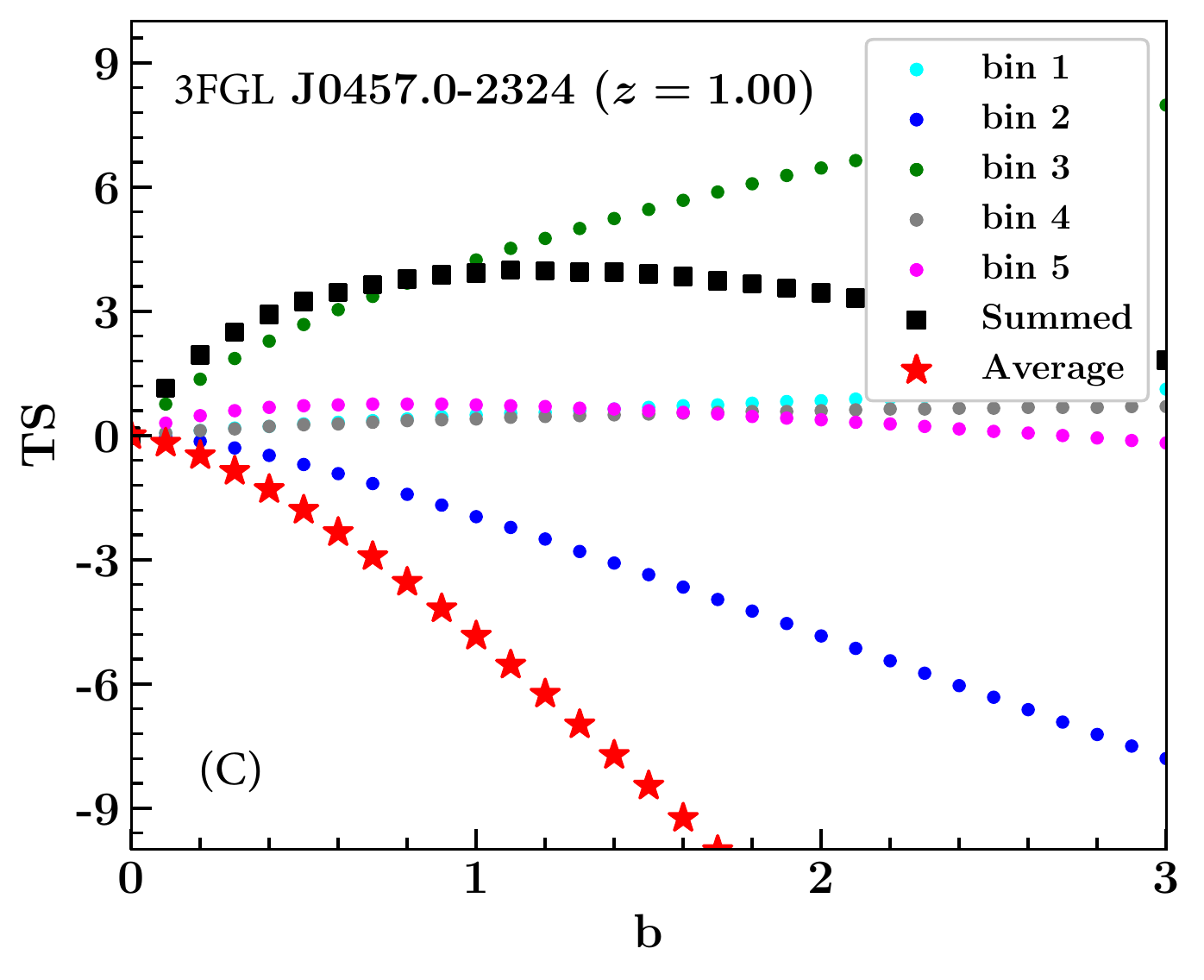} &
  	 \includegraphics[scale=0.55,clip=true,trim=0 0 0
    0]{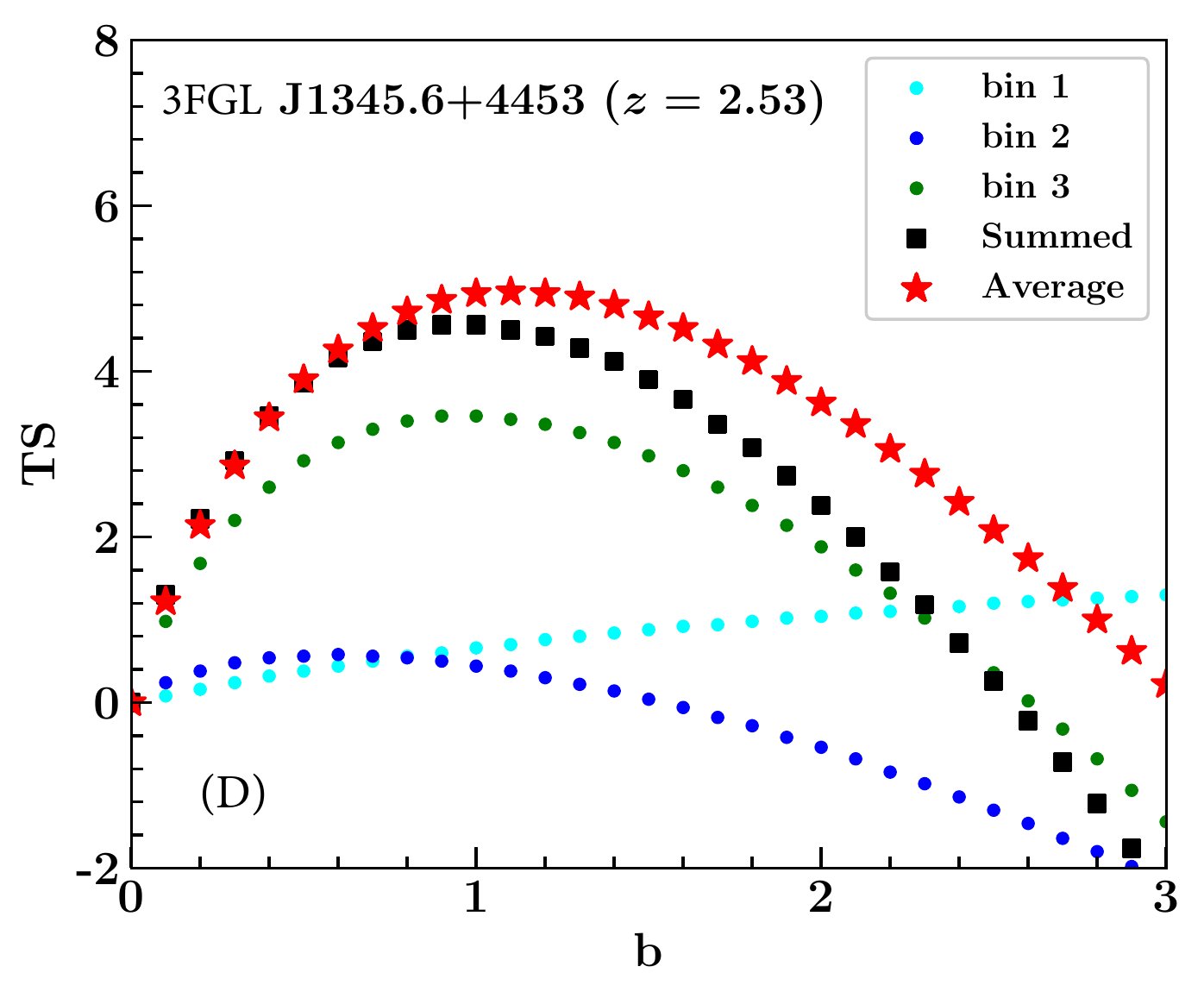} \\
\end{tabular}
  \end{center}
  \caption{{\bf Impact of time-resolved analysis.} Contribution to the $TS_{\rm EBL}$ as a function of scaling
    parameter $b$, adopting the model of \cite{finke10}, for the time resolved (Summed) and time averaged
    (Average) analysis for two BL Lacs (top) and two FSRQs (bottom).
\label{fig:variability}}
\end{figure*}

Blazars are inherently variable objects (at all wavelengths) and
variability may  bias or complicate the measurement of the
EBL attenuation. In order to cope with blazars' variability as much as possible, 
a time-resolved analysis is performed for all sources that are found variable at 
$>10$\,GeV in the recent third catalog of hard sources, 3FHL, \cite{3FHL}.
We rely on the time bins derived by the Bayesian block analysis presented
in 3FHL as these are times when the sources were found to alter their state at $>10$\,GeV,
which is the relevant energy range for detecting the EBL
attenuation. In each time bin, the criteria reported in Table~\ref{tab:tsc} are used to 
determine the best intrinsic spectral model. Because for a given source
time-resolved spectra can be treated as independent observations, their 
contribution to the $TS_{\rm EBL}$ has been summed to the one of the remainder
of the sample. Figure~\ref{fig:ts_vs_b} includes the contribution from variable and non-variable sources.
The level of EBL as determined from the variable sources alone is found to be in good agreement
with the rest of the sample. Figure~\ref{fig:variability} shows the $TS_{\rm EBL}$ as a  function of the $b$ parameter
(for the model of \cite{finke10}) for 4 variable  BL Lacs and FSRQs and how that compares to 
the result of a time-averaged analysis. In general, we find a
time-averaged analysis works well for objects which 
vary primarily in flux, while a time-resolved analysis is required for all
those objects experiencing also spectral variability (see right
versus left plots in Figure~\ref{fig:variability}). Finally, we used the
  Fermi All-sky Variability Analysis tool \cite{fava13,fava17} to
  search for significant residual spectral variability within Bayesian
blocks, but none could be found.

\subsection*{Systematic Uncertainties}
\label{sec:syst}
In order to gauge the systematic uncertainties of this analysis we
have performed the tests reported below:

\begin{itemize}
\item Instead of using a variable maximum energy up to which 
 to fit  the intrinsic spectrum (chosen to be the energy corresponding
 to $\tau_{\gamma\gamma}<0.1$ for the model of \cite{finke10}), we use
 a constant maximum energy of 10\,GeV for all sources. Repeating the
 entire analysis we find $b=1.09\pm0.08$ in agreement with the results presented
in the main text.

\item A similar result as above 
has been obtained using a maximum energy, to measure
  the intrinsic spectrum, defined as that obtained when \tgg$<0.05$
  ($b=1.07\pm0.08$  for the model of \cite{finke10}).

\item We use the IRF bracketing method as described in \cite{lat_perf}.  By
deriving two different sets of IRFs and repeating the entire analysis
we find that the systematic uncertainty in the optical depth $\tau_{\gamma
  \gamma}$ is of the order $\sim$7\,\%.  
\end{itemize}

The results presented in the above sections are fully confirmed and
the systematic uncertainty on the optical depth $\tau_{\gamma \gamma}$ due
to changing the energy threshold to characterize the intrinsic spectrum and IRF are, together, $\lesssim$10\,\%.
A systematic uncertainty of $0.1\times\tau_{\gamma\gamma}$ (added in
quadrature) has been included in the uncertainties reported in
Figures~\ref{fig:horizon} and \ref{fig:taus} and propagated to all
results that use those data.

%
%
\section*{Reconstructing the evolving EBL}
\label{sec_kari}

The optical depth for a $\gamma$ ray of observed energy $E_\gamma$ originating in a source at redshift $z_s$ is related to the evolving number density of EBL photons, $n_{\rm EBL}(\e,z)$, \cite{gould67,Stecker71,Brown73}:
\begin{equation} \label{eq_tau}
    \tau_{\gamma \gamma}(E_\gamma,z_s) = c\int_0^{z_s} \left|\frac{dt}{dz}\right| dz  \int_{-1} ^1(1-\mu)\frac{d\mu}{2} \int^\infty_{ 2m_e^2c^4/\e_{\gamma}(1-\mu) }  \sigma(\e_{\rm EBL},\e_\gamma,\mu) n_{\rm EBL}(\e,z) d\e_{\rm EBL}
\end{equation}
where the rest-frame energy of $\gamma$ rays and EBL photons are
denoted by $\e_\gamma=E_\gamma(1+z_s)$ and $\e_{\rm EBL}=E_{\rm
  EBL}(1+z_s)$ respectively, $\mu = \cos{\theta}$ denotes the angle of
incidence between the two photons, and
$|dt/dz|^{-1}=H_0(1+z)\sqrt{\Omega_M(1+z)^3+\Omega_\Lambda}$, where
the Hubble and the cosmological parameters are
$H_0=70$\,km s$^{-1}$ Mpc$^{-1}$, $\Omega_M$=0.3, and $\Omega_{\lambda}=$0.7. The
cross section for the photon-photon interaction appearing in the last integral in Equation \ref{eq_tau} is
\begin{equation}
  \sigma(\e_{\rm EBL},\e_\gamma,\mu)  = \frac{3\sigma_T}{16}(1-\beta^2)\left[2\beta(\beta^2-2) + (3-\beta^4) \ln{\left( \frac{1+\beta}{1-\beta}\right)} \right],
\end{equation}
with
\begin{equation*}
  \beta = \sqrt{ 1 - \frac{2m_e^2c^4}{ \e_{\rm EBL}\e_\gamma (1-\mu) } }.
\end{equation*}

where $m_ec^2$ is the electron rest mass.
In other words, for a given cosmology, the SED 
and evolution of the EBL uniquely specify the optical
depth at all redshifts. Conversely, we can use the measured optical
depths $\tau_{\gamma\gamma}(E_\gamma,z)$ to reconstruct $n_{\rm EBL}(\e,z)$.

The physical properties of galaxies, such as star-formation rate,
stellar mass and metallicity, are encoded in their SED. Rather than
the EBL, which is accumulated over cosmic time, it is more informative
to study the instantaneous SED of the galaxy population as a whole
i.e., the cosmic emissivity. The buildup of the EBL is related to the volume emissivity $j(\e,z)$ (or equivalently, luminosity density) via:
\begin{equation} \label{eq_n}
  n_{\rm EBL}(\e,z) = (1+z)^3 \int_{z}^\infty  \frac{j(\e^{\prime},\bar{z})}{\e^{\prime}}
  \left|\frac{dt}{d\bar{z}}\right| d\bar{z} ,
\end{equation}
where $\epsilon^{\prime}=\epsilon(1+\bar{z})$ is the rest-frame energy at $\bar{z}$.
The EBL spectral intensity (see Figure~\ref{fig:ebl}) can be found from the number density by
$\epsilon I(\epsilon, z) = \frac{c}{4\pi} \epsilon^2 n_{\rm
  EBL}(\epsilon, z)$.

\subsection*{Model for the Cosmic Emissivity} \label{sec_jnumodel}
Our task is to reconstruct $j(\e,z)$ based on the measured optical depths reported
in Figure~\ref{fig:taus} without making assumptions on galaxy
properties or their stellar population.
We represent $j(\lambda)$ as the sum of several log-normal templates with a fixed peak position:
\begin{equation} \label{eqn:jsum}
j(\lambda) = \sum_i a_i \cdot \exp \left[ -\frac{\left (\log \lambda -
      \log \lambda_i \right)^2 }{2 \sigma^2}   \right] \ \ \ [{\rm erg\cdot s^{-1}cm^{-3}{\Hz}{\rm ^{-1}}}]
\end{equation}
where we fix $\sigma=0.2$, $\lambda_i=[0.17,0.92,2.2,8.0]\,{\rm \upmu
  m}$ and the amplitudes
$a_i$ are left free to vary. We find that four log-normal templates
allow for a sufficiently flexible spectral shape from UV to the mid-IR. A
Lyman-break is imposed by cutting off the spectrum at $\e>13.6$\,eV
where neutral hydrogen becomes opaque. We have chosen the
fixed locations ($\lambda_i$) and width ($\sigma$) of the templates
 such that common features in galaxy SEDs, a flat far-UV continuum and a
4000{\AA}/Balmer break, are easily captured. 
Each template is allowed to evolve independently with redshift based on a
function similar to the SFH parametrization of \cite{Madau2014}
leading to the full expression
\begin{equation} \label{eqn_jnu}
  j(\lambda_i,z) = \sum_i a_i \cdot \exp \left[ -\frac{\left (\log \lambda -
      \log \lambda_i \right)^2 }{2 \sigma^2}   \right] \times
\frac{(1+z)^{b_i}}{1+\left( \frac{1+z}{c_i}\right)^{d_i}}.
\end{equation}
At each of the fixed wavelengths $\lambda_i$, one
parameter controls the amplitude, $a_i$, and three
control the evolution, $b_i$, $c_i$ and $d_i$, yielding a total of $4\times 4=16$
free parameters. The number of optical depth data points is 60.

To explore the sensitivity to different functional forms for the
evolution, we also test the parametrization from \cite{Cole01}:
\begin{eqnarray} \label{eqn_cole}
j(\lambda_i,z) \propto \frac{a_i + b_i z}{1+(z/c_i)^{d_i}}\ ,
\end{eqnarray}
with free parameters $a_i$, $b_i$, $c_i$, and $d_i$ which we display
alongside our main results for the SFH in Figure \ref{fig_sfh_all}.

\subsection*{Markov Chain Monte Carlo: Setup}

We use the MCMC code \texttt{emcee}
\cite{foreman13}, a Python implementation of an affine invariant MCMC
ensemble sampler \cite{GoodmanWeare10}, to constrain the parameters
controlling the emissivity.
The likelihood function is estimated as ${\mathcal L} \propto \exp{(-\chi^2)}$ where $\chi^2$ is given by
\begin{flalign}
\label{eq:chi}
\chi^2 = 
\sum_{i=1}^N \sum_{j=1}^M \frac{ [\tau_{\rm data}(E_i,z_j) - 
    \tau_{\rm model}(\vec{\theta}|E_i,z_j)]^2}{\sigma_{i,j}^2}
\end{flalign}
where there are $N$ energy ($E_i$) bins, $M$ redshift ($z_j$) bins,
$\tau_{\rm data}(E_i,z_j)$ is the measured absorption optical depth
presented in Figure~\ref{fig:taus}, $\tau_{\rm model}(\vec{\theta}|E_i,z_j)$
is the model absorption optical depth with parameters $\vec{\theta}$,
and $\sigma_{i,j}$ is the (statistical plus systematic) uncertainty on the absorption optical depth
measurements.

We choose flat priors on all parameters $\log{a_i} / {\rm (erg s^{-1}
  Mpc^{-3}Hz^{-1})} =[22,32]$, $b_i=[-2,10]$, $c_i=[1,7]$, $d_i=[0,20]$.
We restrict the range of the evolution parameter controlling
the location of the peak (or curvature) to $c_i=[1,7]$ since our
dataset has limited constraining power for shape changes at redshifts
much larger than our sample coverage ($0<z<4$). Note however, that
this does not force the presence of a peak and a turnover as the
function reduces to a power-law $\propto (1+z)^{b_i}$ when $d
\rightarrow 0$.

With the emissivity specified as a function of wavelength and
redshift, we calculate the resulting EBL and optical depth according
to Equations~\ref{eq_tau}--\ref{eq_n} respectively for each
proposed step in the MCMC. Each calculation of $\tau_{\gamma\gamma}$
involves integrating over wavelength, redshift and angle of incidence, but we only require $\tau_{\gamma\gamma}$
at six energies, for every redshift, making it computationally
manageable. Our final results
are based on MCMC chains from 120 walkers exploring the parameter space in 10,000 steps each. This results in 1,140,000 steps after a burn-in of 500 steps for each walker.

\begin{figure}
  \includegraphics[width=0.98\textwidth]{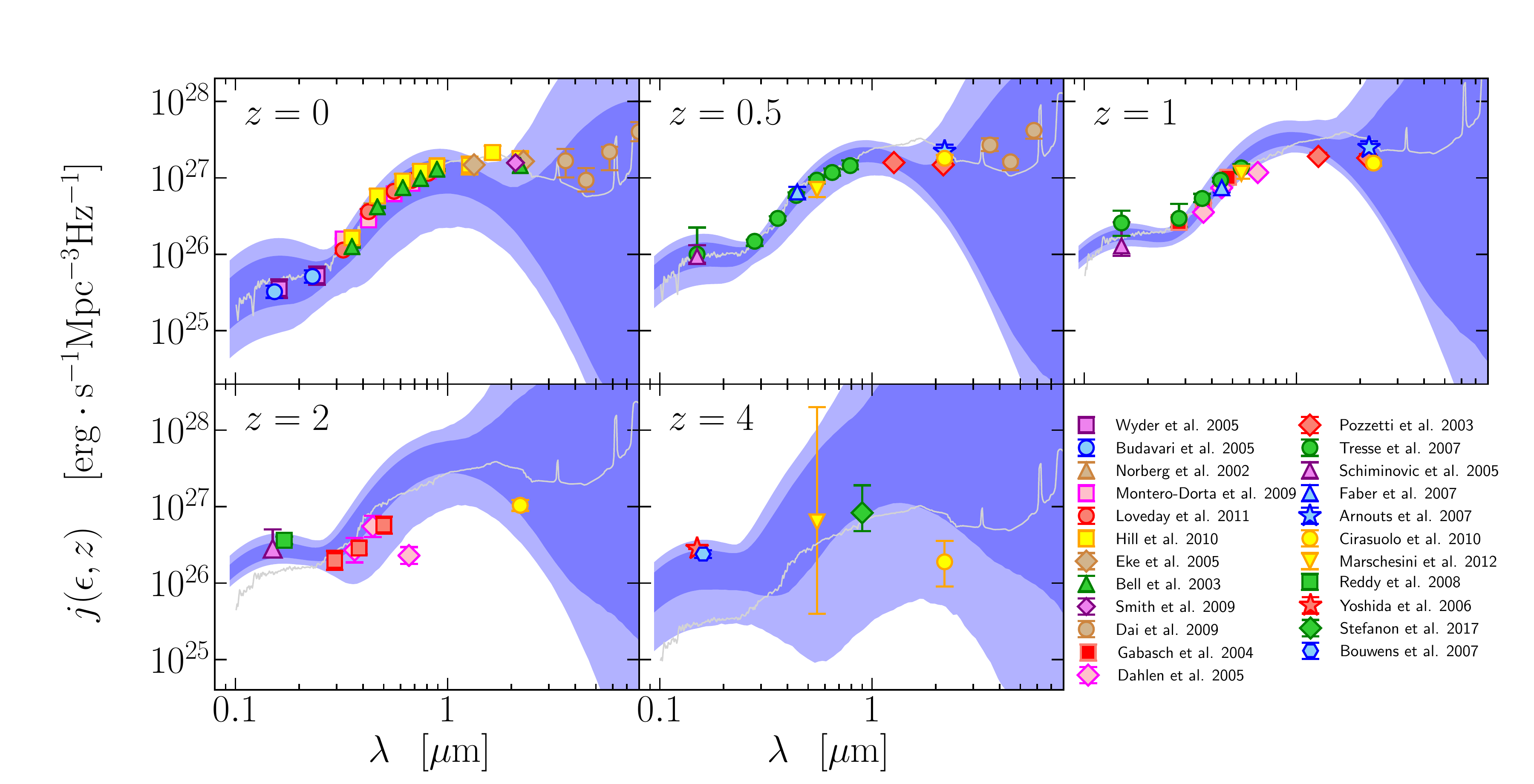}
\caption{{\bf The cosmic emissivity (luminosity density) as a function of
  wavelength in several redshift slices.} The blue shaded regions
  correspond to the 1\,$\sigma$ and 2\,$\sigma$ confidence regions
  resulting from the empirical EBL reconstruction method. The data points are independent measurements from integrated galaxy luminosity functions in the literature. We have not corrected the data for evolution from the redshift displayed (e.g. $z=1$ panel shows measurements at 0.9$<z<1.15$) which may cause some additional scatter. Our results are in general agreement with the galaxy survey data. The gray lines correspond to the EBL model of \cite{dominguez11} where the luminosity density is found to be dominated by a spiral-type galaxy SED template. }
\label{fig_jnu}
\end{figure}

\subsection*{Results and Validation}

In Figure~\ref{fig_jnu} we display the 68\% and 95\% confidence regions for the
 total cosmic emissivity in several redshift
bins. The {\it Fermi}-LAT dataset is tightly constraining at UV,
optical and, at low-$z$, also near-IR wavelengths. The confidence regions get broader
towards mid-IR wavelengths due to the energy range of {\it Fermi} LAT being
limited to $<1$ TeV. Figure ~\ref{fig_jnu} also shows that the {\em
  Fermi}-LAT dataset provides the tightest constraints
around $z\simeq 0.5-1.5$ as the opacity
increases for larger distances traveled. At $z\gsim 2$ we are limited
by the number of bright blazars with substantial emission above
$\sim$10\,GeV in our sample. Comparing our cosmic emissivity with measurements of integrated
galaxy luminosity functions shows that our results are in good overall
agreement across the wavelength range.  This implies that the bulk of the EBL is already accounted for by galaxy surveys. 

We have validated this reconstruction method by creating ten sets of
fake
$\tau(E,z)$ data points in the same energy and redshift bins, and
possessing the same fractional uncertainties, as the original
dataset. The simulated datasets are generated by drawing random sets of
values for the 16 parameters of Equation~\ref{eqn_jnu}, within their assumed
priors, and calculating the optical depths at each energy
and redshift bin. Our reconstruction recovers the fake EBL in all
cases within the derived 2\,$\sigma$ uncertainty region. We see no evidence
for a systematic over- or underestimation of the emissivity at any particular wavelength. The
largest offsets arise at $>1{\rm \upmu m}$ where the dataset is less constraining.

%
%
\subsection*{Comparison with EBL models and data}
\begin{figure}[h!]
\centering
  \includegraphics[width=0.95\textwidth]{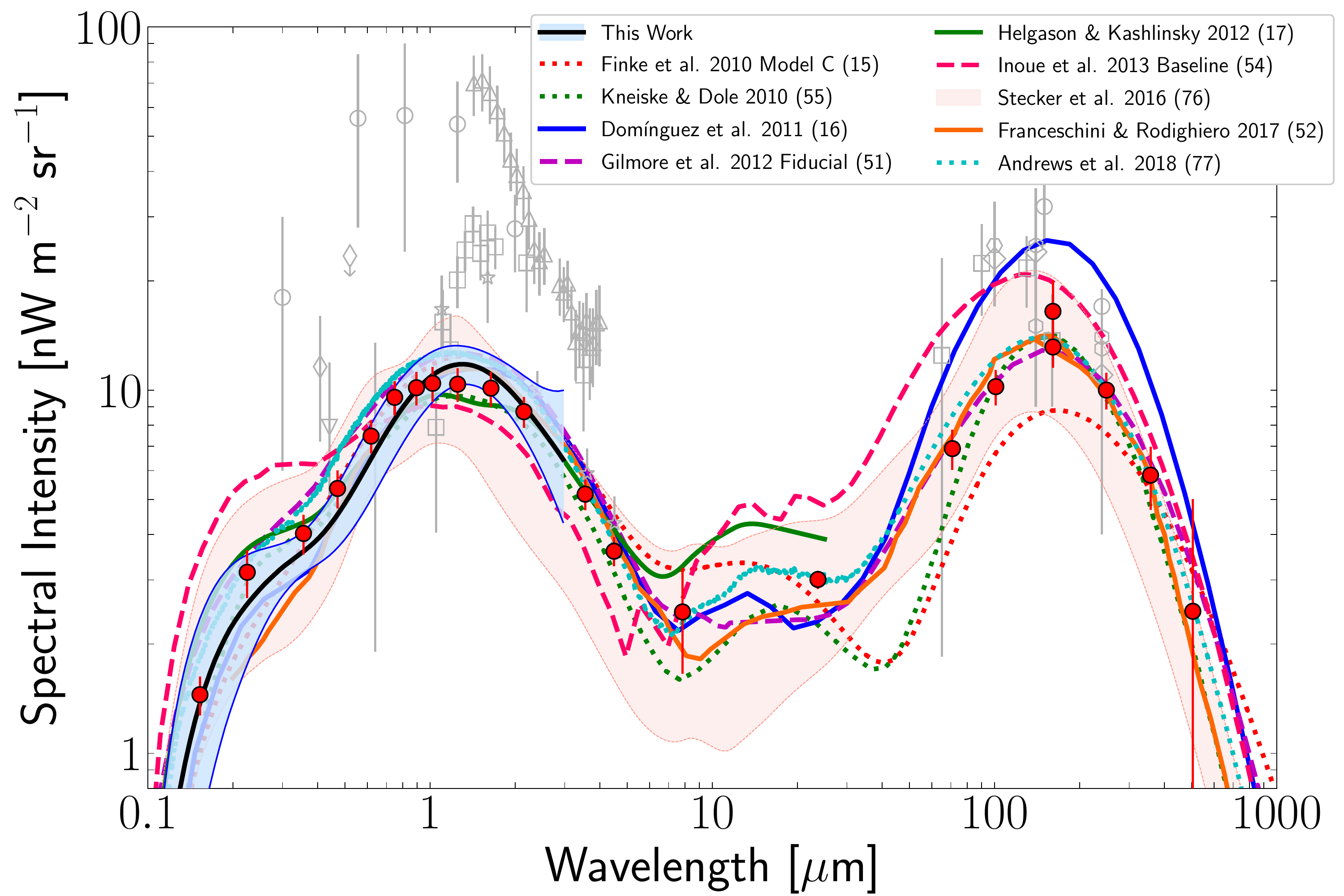}
\caption{{\bf SED of the EBL at $z=0$.} Recovered EBL spectral energy
  distribution at $z=0$ (solid black line) with its $1\sigma$
  uncertainties (shaded blue) in comparison with the some estimates
  from empirical models from ultraviolet to far-IR wavelengths. We
  show some examples for different modeling methodologies:
  observationally motivated (solid lines), physically motivated
  (dotted lines), and theoretically motivated (dashed lines). Our
  uncertainties start to diverge above the near-IR as a consequence of
  the larger uncertainties of our optical-depth data at the larger
  $\gamma$-ray energies. A compilation of data from direct detection
  (open gray symbols) and galaxy counts (filled {red} symbols,\cite{driver16}) is also
  shown. Our spectral intensities match those results from galaxy
  counts leaving little room for substantial contributions from
  sources that have not been detected in deep surveys.}
\label{fig:ebl_models}
\end{figure}

The literature offers different approaches to estimate the evolving
EBL . There are methodologies that are observationally motivated
\cite{dominguez11,helgason12,stecker16,driver16,franceschini17},
physically motivated \cite{kneiske10,finke10,khaire15,andrews18}, and based on semi-analytical models of galaxy formation such as \cite{gilmore12,inoue2013}. Typically these models are constructed in such a way that the lower redshifts and, in general, the optical/near-IR peak are better constrained. Figure~\ref{fig:ebl_models} shows our reconstructed EBL spectral intensities in comparison with some of the models.

The reconstructed EBL follows galaxy counts
\cite{keenan10,voyer11} leaving little room for substantial
contributions not resolved by deep galaxy surveys. This is in conflict
with several direct measurements of EBL (which may be
contaminated by foregrounds, see e.g. \cite{wright00,
  matsumoto05, matsuura17}) and
in tension with some models proposed to explain the anisotropies measured in
diffuse light \cite{zemcov14}. 

Relative to EBL models, in the local Universe, we find that our estimate roughly follows the median of existing models. The models by \cite{finke10}, \cite{kneiske10}, \cite{dominguez11}, \cite{helgason12}, and \cite{franceschini17} reproduce our results quite well. The fiducial model by \cite{gilmore12} tends to follow the upper region of our 1$\sigma$ band. Finally, the baseline model by \cite{inoue2013} provides too much UV and too little near-IR.

The strategy of using the observation of $\gamma$-ray photons to
derive constraints on the background has been used extensively in
recent years. Early attempts, characterized by scarcer $\gamma$-ray
data, only allowed intensity upper limits as a consequence
of assumptions on the intrinsic spectra of blazars
\cite{aharonian06,magic08,meyer12}. These results were
followed by more sophisticated approaches based on more and better
data that allowed the EBL detection and study, both with the LAT
at somewhat lower energies and
thus, larger redshifts \cite{ebl12,dominguez15,armstrong17,desai2017}, and with Imaging Atmospheric Cherenkov
Telescopes (IACTs). {These results from IACTs mostly constrain the local Universe
\cite{abramowski13,biteau15,abdalla17}, although the MAGIC collaboration also probed the EBL at $z\sim 1$ with the detection of two blazars \cite{magic15, magic16}}. 
Notably, our derived EBL at $z=0$ is even closer to the integrated counts
compared to previous $\gamma$-ray derived EBL measurements.

Table~\ref{tab:ebl_intensity} reports the EBL spectral intensities at
several redshifts as displayed in Figure~\ref{fig:ebl}.
Using Equation~\ref{eq_tau}, we can calculate the optical depth curves
as a function of energy and redshift from our reconstructed EBL.
These can be used to correct spectra of cosmological $\gamma$-ray
sources for EBL absorption in order to study physical properties of
the source and are provided \href{https://figshare.com/s/14f943002230d69a4afd}{online}.

\begin{deluxetable}{lcccc}
\tablewidth{0pt}
\tablecaption{Spectral intensity\tablenotemark{a}\ \ ($\lambda I_{\lambda}$) of the EBL as a function of redshift
  and wavelengths as reported in Figure~\ref{fig:ebl}. {The intensities reported in the this table are in comoving
  coordinates. To reproduce the results of Figure~\ref{fig:ebl} (shown
  in physical coordinates) they need to be multiplied by a $(1+z)^3$
  factor. A machine-readable version of this table is provided in Data
  S1 and \href{https://figshare.com/s/14f943002230d69a4afd}{online}.}
\label{tab:ebl_intensity}}
\tablehead{
\colhead{$\lambda$} & \colhead{$z=0$} & \colhead{$z=1$} &
\colhead{$z=2$} & \colhead{$z=3$}   \\
\colhead{($\upmu$m)} & \colhead{(nW m$^{-2}$ sr$^{-1}$)} & \colhead{(nW
  m$^{-2}$ sr$^{-1}$)} & \colhead{(nW m$^{-2}$ sr$^{-1}$)}  & \colhead{(nW m$^{-2}$ sr$^{-1}$)}
}
\startdata
0.102 & $0.12^{+0.12}_{-0.07}$ & $0.41^{+0.08}_{-0.06}$ & $0.36^{+0.11}_{-0.07}$ & $0.21^{+0.07}_{-0.08}$\\ 
0.111 & $0.33^{+0.31}_{-0.20}$ & $0.90^{+0.18}_{-0.13}$ & $0.72^{+0.20}_{-0.14}$ & $0.39^{+0.14}_{-0.15}$\\ 
0.120 & $0.59^{+0.51}_{-0.34}$ & $1.42^{+0.29}_{-0.21}$ & $1.05^{+0.28}_{-0.21}$ & $0.55^{+0.21}_{-0.23}$\\ 
0.130 & $0.90^{+0.68}_{-0.49}$ & $1.92^{+0.38}_{-0.29}$ & $1.33^{+0.36}_{-0.28}$ & $0.68^{+0.28}_{-0.30}$\\ 
0.141 & $1.22^{+0.83}_{-0.63}$ & $2.37^{+0.44}_{-0.37}$ & $1.55^{+0.40}_{-0.34}$ & $0.78^{+0.34}_{-0.35}$\\ 
0.153 & $1.55^{+0.92}_{-0.74}$ & $2.74^{+0.50}_{-0.43}$ & $1.71^{+0.45}_{-0.38}$ & $0.84^{+0.39}_{-0.39}$\\ 
0.166 & $1.87^{+0.98}_{-0.84}$ & $2.99^{+0.56}_{-0.46}$ & $1.80^{+0.47}_{-0.42}$ & $0.87^{+0.43}_{-0.42}$\\ 
0.180 & $2.16^{+0.99}_{-0.87}$ & $3.15^{+0.59}_{-0.48}$ & $1.82^{+0.48}_{-0.44}$ & $0.87^{+0.45}_{-0.42}$\\ 
0.195 & $2.44^{+0.93}_{-0.89}$ & $3.21^{+0.58}_{-0.48}$ & $1.79^{+0.47}_{-0.45}$ & $0.84^{+0.46}_{-0.42}$\\ 
0.212 & $2.68^{+0.87}_{-0.86}$ & $3.19^{+0.55}_{-0.47}$ & $1.72^{+0.46}_{-0.44}$ & $0.80^{+0.45}_{-0.41}$\\ 
0.230 & $2.86^{+0.79}_{-0.79}$ & $3.10^{+0.50}_{-0.44}$ & $1.62^{+0.45}_{-0.42}$ & $0.75^{+0.45}_{-0.38}$\\ 
0.249 & $3.01^{+0.70}_{-0.69}$ & $2.98^{+0.45}_{-0.40}$ & $1.52^{+0.43}_{-0.40}$ & $0.71^{+0.42}_{-0.36}$\\ 
0.270 & $3.12^{+0.60}_{-0.56}$ & $2.85^{+0.41}_{-0.36}$ & $1.44^{+0.40}_{-0.39}$ & $0.68^{+0.39}_{-0.35}$\\ 
0.293 & $3.23^{+0.50}_{-0.48}$ & $2.75^{+0.38}_{-0.33}$ & $1.40^{+0.38}_{-0.39}$ & $0.66^{+0.41}_{-0.36}$\\ 
0.318 & $3.33^{+0.44}_{-0.41}$ & $2.72^{+0.34}_{-0.32}$ & $1.39^{+0.42}_{-0.43}$ & $0.68^{+0.46}_{-0.39}$\\ 
0.345 & $3.46^{+0.41}_{-0.42}$ & $2.77^{+0.38}_{-0.35}$ & $1.45^{+0.52}_{-0.53}$ & $0.70^{+0.56}_{-0.43}$\\ 
0.374 & $3.63^{+0.46}_{-0.48}$ & $2.96^{+0.44}_{-0.43}$ & $1.57^{+0.71}_{-0.65}$ & $0.74^{+0.75}_{-0.48}$\\ 
0.405 & $3.87^{+0.59}_{-0.62}$ & $3.27^{+0.57}_{-0.56}$ & $1.77^{+0.96}_{-0.83}$ & $0.84^{+0.98}_{-0.58}$\\ 
0.440 & $4.21^{+0.75}_{-0.80}$ & $3.73^{+0.72}_{-0.77}$ & $2.04^{+1.28}_{-1.06}$ & $0.99^{+1.26}_{-0.72}$\\ 
0.477 & $4.64^{+0.92}_{-0.98}$ & $4.34^{+0.94}_{-1.01}$ & $2.38^{+1.69}_{-1.33}$ & $1.20^{+1.63}_{-0.92}$\\ 
0.517 & $5.19^{+1.12}_{-1.20}$ & $5.06^{+1.20}_{-1.27}$ & $2.80^{+2.12}_{-1.63}$ & $1.47^{+2.09}_{-1.18}$\\ 
0.561 & $5.84^{+1.32}_{-1.42}$ & $5.91^{+1.48}_{-1.57}$ & $3.27^{+2.58}_{-1.95}$ & $1.78^{+2.66}_{-1.45}$\\ 
0.608 & $6.59^{+1.50}_{-1.65}$ & $6.81^{+1.77}_{-1.93}$ & $3.79^{+3.07}_{-2.28}$ & $2.13^{+3.32}_{-1.75}$\\ 
0.660 & $7.41^{+1.67}_{-1.87}$ & $7.71^{+2.14}_{-2.23}$ & $4.32^{+3.55}_{-2.60}$ & $2.47^{+4.35}_{-2.06}$\\ 
0.716 & $8.25^{+1.82}_{-2.02}$ & $8.56^{+2.50}_{-2.55}$ & $4.88^{+4.01}_{-2.95}$ & $2.88^{+5.38}_{-2.41}$\\ 
0.776 & $9.09^{+1.90}_{-2.13}$ & $9.30^{+2.87}_{-2.84}$ & $5.48^{+4.56}_{-3.35}$ & $3.27^{+6.58}_{-2.72}$\\ 
0.842 & $9.88^{+1.95}_{-2.16}$ & $9.90^{+3.23}_{-3.09}$ & $5.98^{+5.15}_{-3.72}$ & $3.64^{+7.78}_{-3.02}$\\ 
0.913 & $10.60^{+1.96}_{-2.18}$ & $10.32^{+3.56}_{-3.18}$ & $6.52^{+5.88}_{-4.14}$ & $3.94^{+9.86}_{-3.25}$\\ 
0.990 & $11.15^{+1.88}_{-2.06}$ & $10.60^{+3.84}_{-3.36}$ & $6.83^{+6.60}_{-4.40}$ & $4.26^{+12.24}_{-3.52}$\\ 
1.074 & $11.54^{+1.79}_{-1.90}$ & $10.73^{+4.04}_{-3.48}$ & $6.97^{+7.81}_{-4.51}$ & $4.51^{+15.08}_{-3.73}$\\ 
1.164 & $11.79^{+1.65}_{-1.76}$ & $10.64^{+4.23}_{-3.47}$ & $7.04^{+9.00}_{-4.55}$ & $4.65^{+19.68}_{-3.87}$\\ 
1.263 & $11.86^{+1.50}_{-1.58}$ & $10.46^{+4.47}_{-3.65}$ & $6.91^{+11.27}_{-4.47}$ & $4.72^{+28.24}_{-3.94}$\\ 
1.370 & $11.73^{+1.40}_{-1.41}$ & $10.05^{+4.94}_{-3.72}$ & $6.87^{+13.94}_{-4.59}$ & $4.79^{+40.34}_{-4.00}$\\ 
1.485 & $11.50^{+1.28}_{-1.38}$ & $9.56^{+5.80}_{-3.78}$ & $6.81^{+18.49}_{-4.68}$ & $4.81^{+58.87}_{-4.05}$\\ 
1.611 & $11.07^{+1.35}_{-1.35}$ & $9.09^{+6.77}_{-3.86}$ & $6.52^{+25.41}_{-4.61}$ & $4.67^{+84.71}_{-3.96}$\\ 
1.747 & $10.53^{+1.46}_{-1.42}$ & $8.54^{+8.48}_{-3.96}$ & $6.20^{+37.72}_{-4.52}$ & $4.51^{+117.34}_{-3.86}$\\ 
1.895 & $9.94^{+1.63}_{-1.53}$ & $7.97^{+10.94}_{-4.01}$ & $5.77^{+55.07}_{-4.28}$ & $4.38^{+156.29}_{-3.84}$\\ 
2.055 & $9.34^{+1.87}_{-1.70}$ & $7.57^{+14.47}_{-4.07}$ & $5.51^{+76.16}_{-4.22}$ & $4.40^{+213.73}_{-3.90}$\\ 
2.229 & $8.72^{+2.14}_{-1.86}$ & $7.01^{+20.41}_{-4.10}$ & $5.38^{+101.00}_{-4.31}$ & $4.47^{+271.37}_{-4.03}$\\ 
2.417 & $8.16^{+2.55}_{-2.03}$ & $6.70^{+28.53}_{-4.28}$ & $5.24^{+135.49}_{-4.33}$ & $4.51^{+341.38}_{-4.11}$\\ 
2.621 & $7.57^{+3.20}_{-2.16}$ & $6.31^{+39.23}_{-4.30}$ & $5.28^{+174.14}_{-4.50}$ & $4.72^{+415.24}_{-4.37}$\\ 
2.843 & $7.05^{+4.08}_{-2.33}$ & $5.95^{+52.73}_{-4.31}$ & $5.38^{+223.92}_{-4.71}$ & $4.86^{+494.98}_{-4.55}$\\ 
3.083 & $6.54^{+5.51}_{-2.48}$ & \nodata & \nodata & \nodata \\ 
3.344 & $6.15^{+7.44}_{-2.69}$ & \nodata & \nodata & \nodata \\ 
3.626 & $5.85^{+9.76}_{-2.91}$ & \nodata & \nodata & \nodata \\ 
3.933 & $5.46^{+13.33}_{-3.03}$ & \nodata & \nodata & \nodata \\ 
4.265 & $5.23^{+17.70}_{-3.21}$ & \nodata & \nodata & \nodata \\ 
4.626 & $5.11^{+23.25}_{-3.44}$ & \nodata & \nodata & \nodata \\ 
5.017 & $5.07^{+30.88}_{-3.70}$ & \nodata & \nodata & \nodata \\ 
\enddata
\end{deluxetable}

\clearpage

\subsection*{Implications for the high-$z$ Universe } \label{sec_highz}

\begin{figure}
\centering
  \includegraphics[width=0.78\textwidth]{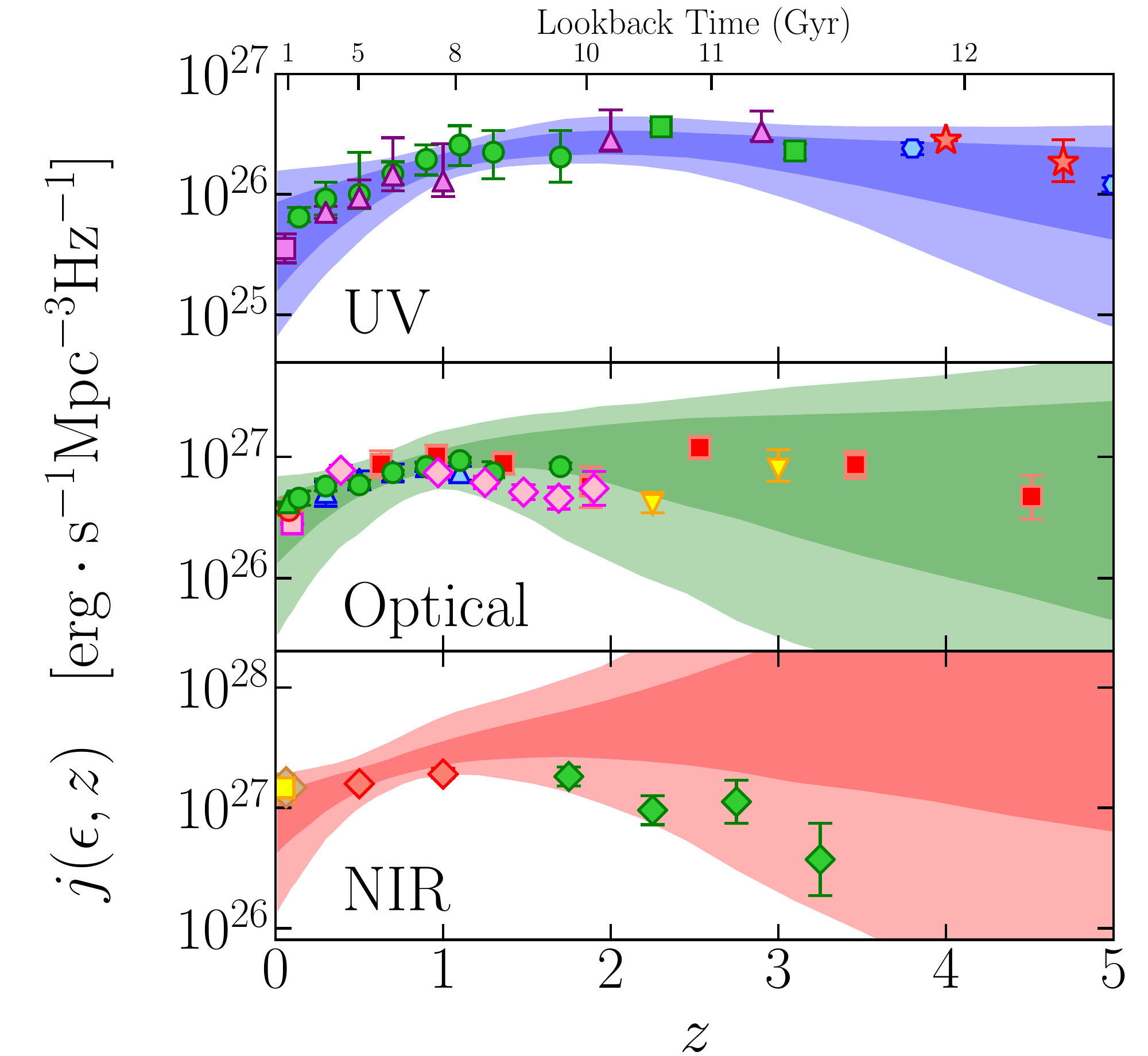}
\caption{\label{fig_ld} {\bf Evolution of the cosmic emissivity.} {The evolution of the cosmic emissivity at UV
  (0.16\,\mic), optical (0.45\,\mic) and NIR (1.6\,\mic), panels A, B
  and C respectively.} The shaded regions show the 1$\sigma$ and 2$\sigma$
  confidence regions resulting from the empirical EBL reconstruction model.
The data points shown have rest-frame wavelengths in the range 0.15-0.17\mic, 0.42-0.48\mic\ and 1.25-1.27\mic\ in the UV, optical, and NIR panels respectively. Colors and symbols follow the same scheme as in Figure \ref{fig_jnu}. }
\end{figure}

In principle, there is always some constraining power beyond the
maximum redshift of the sample of $\gamma$-ray sources. This is because the $\gamma$
rays coming from $z_{\rm max}$
start interacting with EBL photons which were built up at still earlier
times and the rate of the interactions is related to $n_{\rm
    EBL}\propto (1+z)^3$. Focusing on the UV, which is important for
  cosmic re-ionization, Figure \ref{fig_ld} suggests
  rather minimal UV emissivity at $z>4$ with respect to measurements from
  Lyman-break galaxy surveys. However, it is possible that the confidence regions at these redshifts
 may be artificially narrow due to the lack of
  flexibility in the parameterized shape of the evolution. In order to test the robustness of the constraints at
  high-$z$ we have re-run the MCMC and included an additional
  term in Equation~\ref{eqn_jnu}:
\begin{equation}
j_{\rm high-z}(\lambda,z) = a_{\rm high-z} \exp{\left[ -\frac{( z -
      z_0 )^2}{2\sigma^2}
  \right]} \left( \frac{\lambda}{0.17\mathrm{\upmu m}}\right)^{-0.5}
\end{equation}
centered at $z_0=6$ with $\sigma=0.08$. We find that, while
this reveals a relatively unconstrained lower limit for the UV
emissivity, the upper limit remains robust at $\sim 3.2(5.3) \times
10^{26}$ erg s$^{-1}$ Mpc$^{-3}$ Hz$^{-1}$, 1$\sigma$(2$\sigma$), at
$z=5-6$. In Figure \ref{fig_Mcut}, we compare this value with the
integrated UV luminosity functions from the Hubble Frontier
Fields (HFF) program, which targets extremely faint galaxies behind strong
gravitational lenses, reaching $M_{\rm AB}\sim
-13$. Some HFF analyses have found evidence for a turnover
in the steep faint-end of the luminosity function (LF) \cite{Bouwens17,Atek18}, whereas
others do not see such a feature \cite{Livermore17,Ishigaki18}. The
conflicting results at $M_{\rm AB} \gsim -15$ could be due to
uncertainties in the magnification factor determined by lens
models. 

Our constraints limit how far a steep faint-end slope can be
extrapolated. In fact, Figure \ref{fig_Mcut} shows that the emissivities from the integrated UV luminosity
functions are already close to our derived upper limits,
but are all compatible within 2\,$\sigma$. They favor a turnover of the LF at $M_{\rm AB}\sim
-14$ in agreement with \cite{Bouwens17} and \cite{Atek18}. The UV
emissivity implied by \cite{Ishigaki18} for example (with no turnover), would reach the
2$\sigma$ upper limit if extrapolated to $M_{\rm AB}\sim -10$.

In Figure \ref{fig_Mcut}, we also show the UV emissivity necessary to
sustain a reionized Universe at $z=6$. The required emissivity  (at
0.15\,${\rm \upmu m}$) can be shown to be \cite{Madau99}:
\begin{equation}
  j_{\rm UV} = 2.5 \times 10^{26} \hspace{5pt} \epsilon_{53}^{-1}  \left(
    \frac{1+z}{6} \right)^3  \left( \frac{\Omega_bh_{70}^2}{0.0461}
  \right)^2 \left( \frac{C/f_{\rm esc}}{30} \right)^2 \hspace{20pt}
  {\rm erg~s^{-1}Mpc^{-3}Hz^{-1} }.
\end{equation}
Here, $\Omega_b$ is the cosmic baryon density, $h_{70}$ is Hubble
  parameter in units of 70\,${\rm km~s^{-1}Mpc^{-1}}$, $C$ is the
  clumping factor of ionized hydrogen and $f_{\rm esc}$ is the mean escape
  fraction of ionizing photons. The parameter $\epsilon_{53}$ is the number of Lyman continuum photons per
unit of forming stellar mass in units of $10^{53}$ photons $\cdot {\rm
  s^{-1} (M_{\odot}\cdot yr^{-1})^{-1}}$. For this we follow
\cite{Finkelstein12} exploring values of $\epsilon_{53}$ based on
stellar population synthesis models assuming a Salpeter IMF and a
constant star formation rate. The width of the grey regions in
Figure \ref{fig_Mcut} correspond to the range $0.9<\epsilon_{53}<1.4$
when the metallicity is varied from $0.02Z_\odot$ to $1.0Z_\odot$
(where $Z_\odot$ is the solar metallicity).
We display the emissivity for a reasonable assumption of $C/f_{\rm
  esc} = 30$, showing that our constraints accommodate a
scenario in which the Universe is reionized at $z=6$.

Our constraints at $z>4$ come almost
entirely from GRB~080916C which provides a strong upper limit to the
optical depth at $z=4.35$ whereas the blazar sample alone ($z<3.1$) has
lower constraining power.
This is a benefit of detecting more high-$z$ $\gamma$-ray
sources as probes of the epoch of re-ionization \cite{Kashlinsky05}.

\section*{The Star-Formation History}
\label{sec:SFH}

We derive the SFH from our constraints on the far-UV emissivity in a
similar manner to galaxy surveys that measure the rest-frame UV emission
\cite{Schiminovich2005,Bouwens14b,McLure13,Ellis13,finkelstein2015}. The
conversion into star-formation rate (SFR) requires two assumptions: i) the amount of UV
emission expected per unit SFR, $\mathcal{K}_{\rm UV}$, which is dictated
by the initial mass function (IMF) of choice, and ii) the mean dust extinction within the host
galaxies, $A_V$, since photons become a part of the EBL only if they escape their
progenitor galaxies.
For the former quantity, we assume $\mathcal{K}_{\rm UV} = 7.25\times
10^{-29} {\rm M_\odot~yr^{-1}~erg^{-1}~s~Hz}$ which is consistent with
a Chabrier IMF \cite{chabrier03}. Our results on the SFH can be re-scaled by constant
factor of 1.6 to represent a Salpeter IMF \cite{salpeter95}.

\begin{figure*}[ht!]
  \centering
  	 \includegraphics[scale=0.74,clip=true,trim=0 0 0 0]{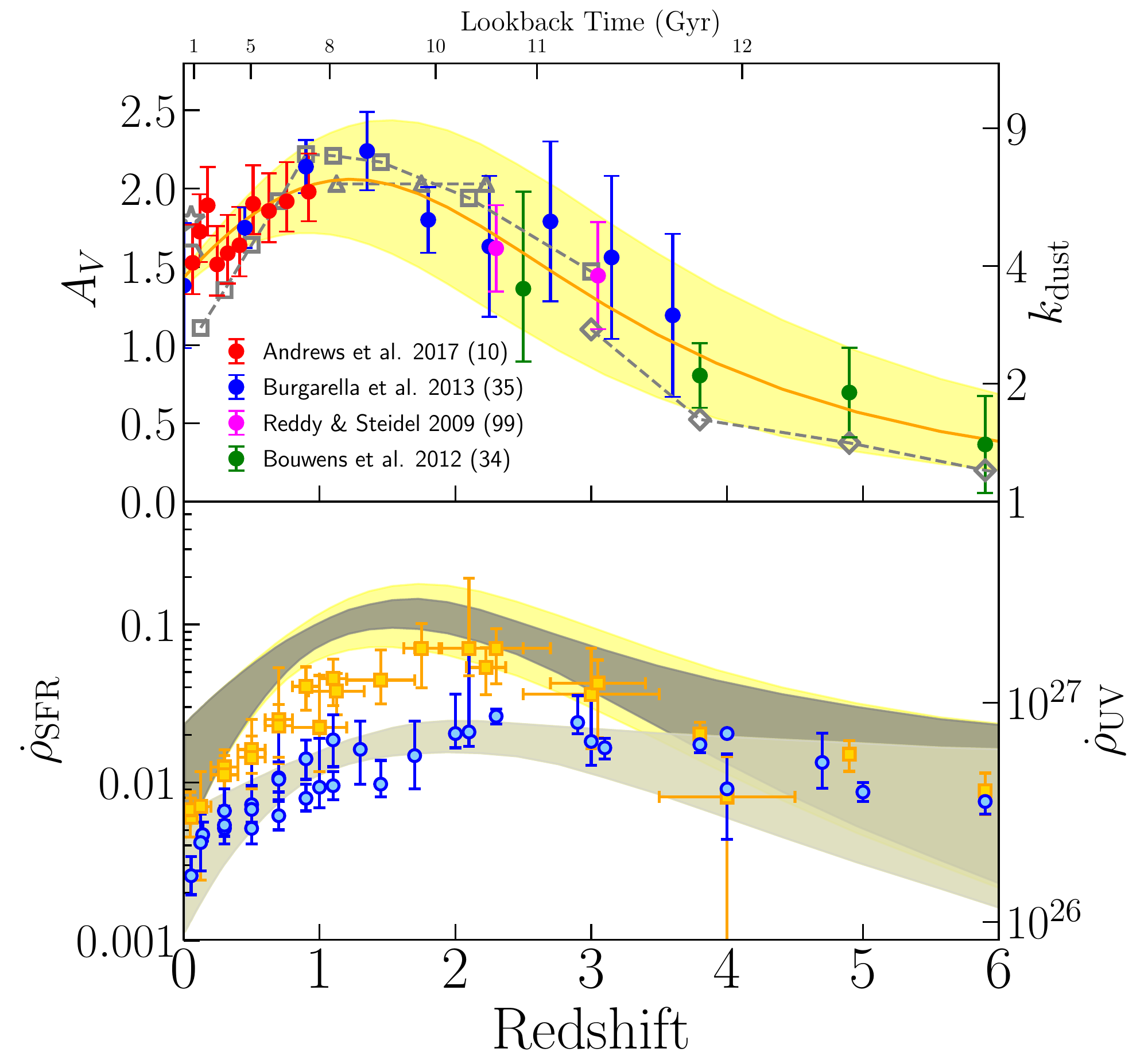} \\
  \caption{{\bf The effects of dust extinction on the derived SFH}.
    {\it Panel A:} The mean dust extinction as a function
    of redshift. The solid line is our best fit (see Equation
    \ref{eq:av}) with uncertainty shown as yellow region. Data points
    used for the fit (filled
    circles)  are from
    \cite{Andrews17,Burgarella13,Reddy2009,Bouwens2012}. Also shown as gray connected points are estimates from other
    references without published uncertainties: \cite{Wyder2005} star, \cite{Robotham2011} hexagon,
    \cite{Dahlen2007} triangles, \cite{Cucciati12} squares,
    \cite{Bouwens16} diamonds.  Right vertical axis shows the
    multiplicative factor $k_{\rm
      dust}=10^{0.4A_V}$. {\it Panel B:} The SFH corrected (dark gray) and
    uncorrected (light gray) for dust extinction (in ${\rm M_\odot~yr^{-1}Mpc^{-3}}$). Yellow region
    includes the systematic uncertainty from the dust correction 
      which has been added
    in quadrature to the statistical uncertainties. The
    data points show the corrected (orange) and uncorrected (blue) SFH
    from the compilation of \cite{Madau2014} with the addition of
    \cite{Yoshida2006} and \cite{Tresse2007}. Right vertical axis
    shows the UV emissivity in units of ${\rm erg~s^{-1}Mpc^{-3}Hz^{-1}}$.
\label{fig_dust}}
\end{figure*}

For the dust extinction correction, we rely on measured values of the mean
$A_V$ from the literature and fit its evolution with redshift using
the following parametrization: $A_V\propto \frac{(1+z)^f}{1+\left( \frac{1+z}{c}\right)^d}$. The result is shown in Figure
\ref{fig_dust}. The measured values of $A_V$ are based on different
methods. For instance, these come from: measured UV continuum
slopes \cite{Bouwens14b,Bouwens16}, stellar population synthesis SED
fitting \cite{Cucciati12,Andrews17} and comparison of the
integrated UV and IR luminosity functions
\cite{Takeuchi05,Burgarella13}. We choose to use only those data  
that are measured from a large sample where robust uncertainty
estimation is provided. Studies that assume or estimate values of
$A_V$ do not contribute to the fit but are shown in Figure
\ref{fig_dust} for reference. We obtain the evolving extinction:
\begin{equation}
\label{eq:av}
  A_V(z) = (1.49\pm 0.07)\frac{(1+z)^{(0.64\pm 0.19)}}{1 +
    [(1+z)/(3.40\pm 0.44)]^{(3.54\pm 0.47)}}.
\end{equation}
The SFH is then calculated as:
\begin{equation}
  {\rm \rho_{\star}}(z)= j_{\rm UV}(z) \cdot  \mathcal{K}_{\rm UV} \cdot 10^{0.4A_V}
\end{equation}
where $j_{\rm UV}(z)$ is our reconstructed emissivity at 1600\AA. 

\begin{figure*}[ht!]
  \begin{center}
  \begin{tabular}{l}
\hspace{-1.0cm}
  	 \includegraphics[scale=0.83,clip=true,trim=0 0 0 0]{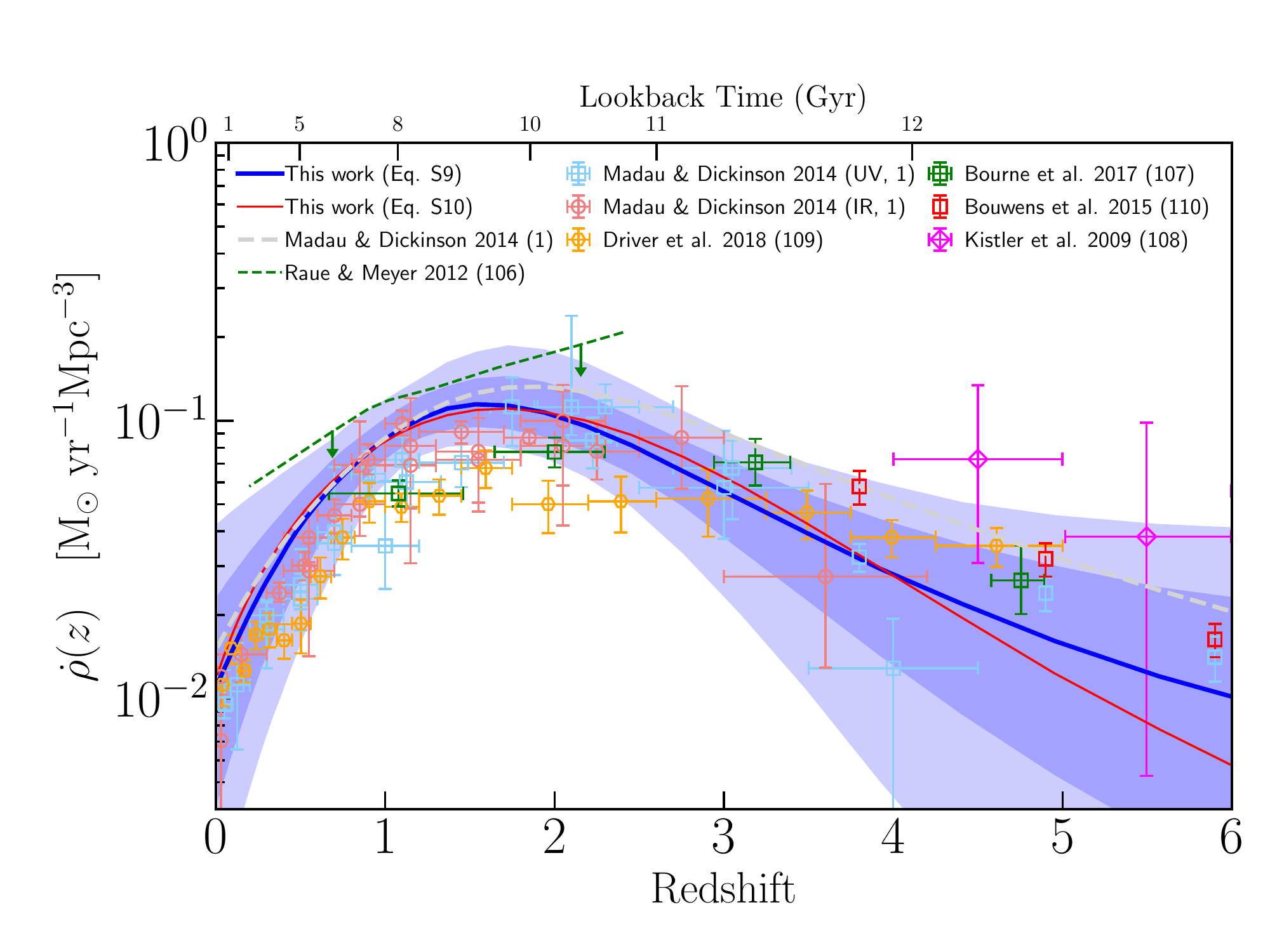}
\end{tabular}
  \end{center}
  \caption{{\bf The star-formation history of the Universe.} 
Results for the SFH compared with data from the literature. The blue and red lines compare
    the median SFH resulting from our EBL reconstruction using the
    evolution parameterization in Eq.~~\ref{eqn_jnu}  from
    \cite{Madau2014}
versus the parametrization in Eq. ~\ref{eqn_cole} from \cite{Cole01}.
The blue regions show the 1$\sigma$ and
    2$\sigma$ confidence intervals {for the  EBL reconstruction model}.
    The dashed gray line shows the fit from \cite{Madau2014} and
    dashed green line are upper limits from $\gamma$-ray data derived
    by \cite{raue12} where they assume a Chabrier IMF and $\beta = 0.3$.   Data points are from the
    compilation of \cite{Madau2014} with the addition of data from 
    \cite{Bourne2017,Kistler2009,Driver2017, Bouwens2015}. The data have been corrected for variations in
    the adopted IMF to $\mathcal{K}_{\rm UV}=7.25\times
    10^{-29}$ consistent with a Chabrier IMF (see text).
\label{fig_sfh_all}}
\end{figure*}

The confidence regions
for the cosmic SFH are shown in Figure \ref{fig:sfr} in the main paper
along with data points from UV-derived measurements \cite{Madau2014}. We also display the same result
in Figure \ref{fig_sfh_all} showing data from various studies using
different tracers of SFR, including limits from $\gamma$-ray
constraints of the EBL \cite{raue12}. At low and intermediate redshifts, our
results are in good agreement with (albeit a little bit above) independent measurements from
galaxy surveys. At $z>3$, our results are in agreement, within the
uncertainties, but favor a rather low SFH. As discussed in the previous
subsection, this is primarily driven by GRB~080916C. More
importantly, because the SFH derived from $\gamma$-ray absorption
complements traditional methods that probe the SFH from sources
resolved in surveys, our results imply that the bulk of star formation
across cosmic time is already accounted for by surveys.

%
%

\subsection*{Stellar Population Model Method}
\label{sec:justin}
The cosmic SFH, $\dot{\rho}(z)$,
 is the starting
point in the EBL model of \cite{razzaque09,finke10} making it a useful model for further
  exploration of the
parameter space that is made possible by the $\g$-ray optical depth data. The model
assumes that stars emit as blackbodies, with their temperatures,
luminosities, and time evolution determined from formulae given by
\cite{eggleton89}. The radiation emitted by stars is convolved with
an IMF and star-formation rate density
parameterization to get the luminosity density $j(\e; z)$.  The
fraction of light that escapes dust extinction ($f_{\rm esc,dust}$) is
based on the extinction curve from \cite{driver08}, which was derived from a
fit to the luminosity density data in the local Universe. We let the dust extinction evolve according to Equation \ref{eq:av}.
The infrared portion of the
EBL is computed assuming that all the energy absorbed by dust is
re-radiated in the infrared.  The SFH and IMF model parameters were
chosen to reproduce the luminosity density data available at the
time. Once $j(\e; z)$ is calculated, the EBL number
  density and $\g$-ray absorption optical depth are computed from
  Equations~\ref{eq_tau}--\ref{eq_n} above.

 Using the methodology of
\cite{finke10}, we have performed an MCMC fit to the
$\g$-ray optical depth data.  We parameterize the SFH and let the parameters vary,
calculating the resulting EBL and optical depths in each step.
A similar MCMC model fit, but limited to $z\geq2$, was done by \cite{gong13} to the earlier
EBL absorption data from \cite{ebl12}. We use the standard
parameterization for the SFH (Equation
\ref{eqn_jnu}), but also consider evolution according to Equation~\ref{eqn_cole}.
The SFH result from our MCMC fits, reported in Figure
\ref{fig:sfr} (as the green confidence region), are consistent
with the SFH used for the ``model C'' of \cite{finke10}, which relied
on  the \cite{Cole01} parametrization with free parameters given
by \cite{hopkins06},
at all values of $z$, but the confidence
interval is particularly narrow up to $z\leq2.5$. 
Table~\ref{tab:sfh_values} reports the values of the SFH obtained from
both methods as displayed in Figure~\ref{fig:sfr}.

\begin{deluxetable}{lcc}
\tablewidth{0pt}
\tablecaption{The Cosmic star-formation history as reported in
  Figure~\ref{fig:sfr}, also available \href{https://figshare.com/s/14f943002230d69a4afd}{online}.
\label{tab:sfh_values}}
\tablehead{
\colhead{$z$} & \colhead{Physical EBL model} & \colhead{EBL Reconstruction} \\
\colhead{} & \colhead{($10^{-2}$M$_\odot$\,yr$^{-1}$\,Mpc$^{-3}$)} & \colhead{($10^{-2}$M$_\odot$\,yr$^{-1}$\,Mpc$^{-3}$)}
}
\startdata

0.0 & $0.8^{+0.7}_{-0.3}$ & $1.2^{+1.3}_{-0.7}$ \\ 
0.1 & $1.1^{+0.8}_{-0.4}$ & $1.6^{+1.4}_{-0.9}$ \\ 
0.2 & $1.6^{+0.9}_{-0.5}$ & $2.1^{+1.4}_{-1.1}$ \\ 
0.3 & $2.0^{+0.9}_{-0.5}$ & $2.7^{+1.4}_{-1.2}$ \\ 
0.4 & $2.6^{+0.8}_{-0.6}$ & $3.4^{+1.4}_{-1.4}$ \\ 
0.5 & $3.3^{+0.8}_{-0.6}$ & $4.2^{+1.4}_{-1.5}$ \\ 
0.6 & $4.1^{+0.8}_{-0.7}$ & $5.0^{+1.4}_{-1.5}$ \\ 
0.8 & $6.0^{+1.0}_{-0.9}$ & $6.8^{+1.5}_{-1.4}$ \\ 
1.0 & $8.2^{+1.6}_{-1.4}$ & $8.5^{+1.6}_{-1.3}$ \\ 
1.2 & $10.7^{+2.0}_{-2.2}$ & $10.0^{+2.1}_{-1.5}$ \\ 
1.4 & $12.9^{+2.5}_{-2.8}$ & $11.0^{+2.3}_{-1.8}$ \\ 
1.6 & $14.5^{+2.3}_{-2.9}$ & $11.3^{+2.9}_{-2.0}$ \\ 
1.8 & $15.0^{+2.3}_{-2.5}$ & $11.0^{+3.1}_{-2.0}$ \\ 
2.0 & $14.7^{+2.4}_{-2.4}$ & $10.4^{+3.0}_{-1.9}$ \\ 
2.5 & $11.6^{+3.2}_{-2.9}$ & $7.9^{+2.2}_{-1.7}$ \\ 
3.0 & $7.6^{+3.3}_{-2.4}$ & $5.6^{+1.8}_{-1.8}$ \\ 
3.5 & $4.8^{+2.6}_{-1.8}$ & $4.0^{+1.5}_{-1.7}$ \\ 
4.0 & $3.1^{+2.0}_{-1.3}$ & $2.8^{+1.5}_{-1.5}$ \\ 
4.5 & $2.0^{+1.6}_{-0.9}$ & $2.1^{+1.4}_{-1.3}$ \\ 
5.0 & $1.3^{+1.3}_{-0.7}$ & $1.6^{+1.4}_{-1.1}$ \\ 
5.5 & $0.9^{+1.1}_{-0.5}$ & $1.2^{+1.3}_{-0.9}$ \\ 
6.0 & $0.6^{+1.0}_{-0.3}$ & $1.0^{+1.3}_{-0.8}$ \\ 
\enddata
\end{deluxetable}

\subsection*{The contribution of Active Galactic Nuclei}
\label{sec:agn}

\begin{figure}
\centering
  \includegraphics[width=0.80\textwidth]{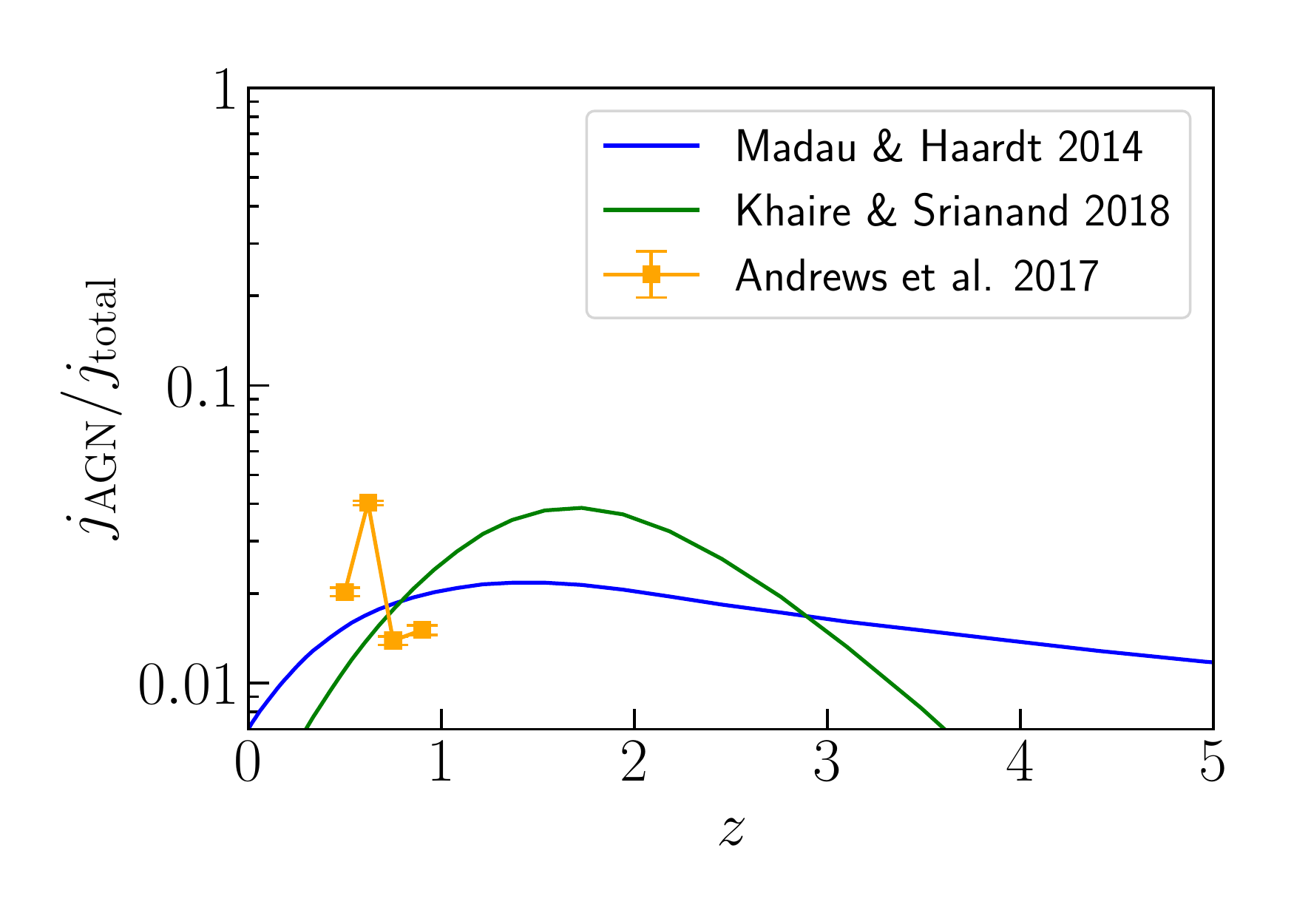}
\caption{Ratio of emissivities of AGN vs total at 0.16\,$\upmu$m. The
  $j(\e,z)_{\rm total}$ comes from the median UV emissivity derived from
  the optical depth data. The AGN emissivity is taken from the
  empirical fit of integrated quasar luminosity functions (converted
  to $0.16{\rm \upmu m}$) provided by \cite{Madau2015} and
  \cite{Khaire2018}, plotted in blue and green respectively.}
\label{fig:agn}
\end{figure}

The methods employed here to derive the SFH of the Universe
rest on the assumption that most of the EBL is the product of stellar
emission. While this is almost certainly true in the IR and optical bands,
there could be a non-negligible contribution of active galactic nuclei (AGN) to the
global UV background \cite{dominguez11}.
We estimate this contribution by considering
measurements of the integrated quasar luminosity function. A fitting
formula for the resulting emissivity as a function of redshift was
provided by \cite{Madau2015} at a rest-frame wavelength of 912\AA. We
convert this to 0.16\,$\upmu$m (the wavelength used to estimate our SFH) adopting the
same power law spectrum $\propto \lambda^{-0.61}$ \cite{Lusso2015}
and show the ratio of the
AGN to total (as estimated in this work) emissivities at  0.16\,$\upmu$m
in Figure~\ref{fig:agn}.
This shows that the contribution from known AGN is 
no more than a $few$  percent. This is in agreement with
the recent estimate of the AGN contribution to the
EBL \cite{andrews18}.

\begin{thebibliography}{10}
\bibitem{note1} 
We adopted the following values for the Hubble and
  cosmological parameters: $H_{0}$=70\,km
  s$^{-1}$ Mpc$^{-1}$,  $\Omega_M$=0.3, and  $\Omega_{\Lambda}$=0.7.
\end{thebibliography}
\end{document}